\documentclass[twocolumn,superscriptaddress,prb,amsmath,amssymb,showpacs,lengthcheck]{revtex4-1} 

\usepackage[T1]{fontenc}
\usepackage{graphicx}
\usepackage{color}
 \usepackage[normalem]{ulem}

\usepackage{braket}
\newcommand\ret{r}
\newcommand\pdiff{\frac{\partial}{\partial t}}

\begin{document}

\title{Interplay of structural design and interaction processes in tunnel-injection semiconductor lasers}

\author{S. Michael}
\affiliation{Institute for Theoretical Physics, University of Bremen, 28359 Bremen, Germany}
\author{M. Lorke}
\affiliation{Institute for Theoretical Physics, University of Bremen, 28359 Bremen, Germany}
\author{M. Cepok}
\affiliation{Institute for Theoretical Physics, University of Bremen, 28359 Bremen, Germany}
\author{C. Carmesin}
\affiliation{Institute for Theoretical Physics, University of Bremen, 28359 Bremen, Germany}
\author{F. Jahnke}
\affiliation{Institute for Theoretical Physics, University of Bremen, 28359 Bremen, Germany}

\begin{abstract}
Tunnel-injection lasers promise various advantages in comparison to conventional laser designs. In this paper, the physics of the tunnel injection process is studied within a microscopic theory in order to clarify 
design requirements for laser structures based on quantum dots as active material and an injector quantum well providing excited charge carriers. 
We analyze how the electronic states of the injector quantum well 
and quantum dot levels should be aligned and in which way their coupling through the tunnel-injection barrier 
should be adjusted for optimal carrier injection rates into the quantum-dot ground state used for the laser transition. 
Our description of the tunnel-injection process combines two main ingredients: the tunnel coupling of the wave functions 
as well as the phonon- and Coulomb-assisted transition rates. For this purpose, material-realistic electronic state calculations 
for the coupled system of injector quantum well, tunnel barrier, and quantum dots are combined with a many-body theory for the 
carrier scattering processes. We find that the often assumed longitudinal-optical-phonon resonance condition for the level alignment has practically 
no influence on the injection rate of carriers into the quantum dot states. The structural design should provide optimal hybridization 
of the injector quantum well states with excited quantum dot states.
\end{abstract}

\maketitle

\section{Introduction\label{Introduction}}
In conventional semiconductor lasers, the modulation response is limited by the nonlinearity of the differential gain. 
This nonlinearity originates from several effects, the most important ones being spectral hole burning and carrier heating. 
The latter can be mitigated in semiconductor lasers with quantum wells (QWs) 
as active material by injection of cold carriers through a tunnel-injection (TI) barrier.\cite{zhang19970} Hereby, the 
temperature stability and modulation speed have been improved. TI structures are also promising to overcome current limitations 
of quantum-dot (QD) devices \cite{bhattacharya2003carrier,fathpour2005high} and progress in growth and optical characterization 
of QD-TI devices \cite{bhattacharya2002tunnel,mi2005high,mi2006growth,lee2011characteristics,bhowmick2014high,podemski2006tunnel,skek2007experimental,syperek2010time,syperek2012influence,rudno2012electronic} 
was made.

In a QD laser, the pump process typically generates excited carriers in the delocalized states, while the QD ground state is used for the carrier recombination into the laser mode. 
The capture of carriers from extended into localized states is assisted by carrier-phonon \cite{Inoshita:92,Singh:98,Seebeck:05} and carrier-carrier Coulomb scattering processes.
\cite{Bockelmann:92,Uskov:97,Magnusdottir:03,Nielsen:04} The design with an injector QW separated by a thin TI barrier from QDs (see Fig.~\ref{fig1}) can lead to 
more efficient capture of excited carriers from extended into localized states, as the energy difference between injector QW states and QD states can be engineered to 
lower values in comparison to energy differences occurring in dot-in-a-well or dot-on-wetting-layer structures. Furthermore, the TI design contributes to a suppression 
of detrimental hot carrier effects.\cite{mi2005high} In recent experiments, improvements of GaAs-QD based high-power lasers due to the TI scheme has been demonstrated 
\cite{pavelescu2009high} and ultra-fast gain recovery has been achieved.\cite{pulka2012ultrafast} 
To benefit from the advantages of the TI design in telecommunication applications, InAs QDs within an InGaAlAs barrier lattice matched to the InP substrate are attractive 
due to their emission around the 1.55 $\mu$m wavelength and their reduced size inhomogeneity. 
\cite{marynski2013electronic,bhowmick2014high,banyoudeh2015high,syperek2016exciton,bauer2018growth,rudno2018carrier,rudno2018control}
Further investigations of the InAs/InP material system include  studies of the behavior of the QD ground state under the influence of varying QW parameters,
as well as the carrier dynamics in TI structures at cryogenic temperatures.\cite{syperek2018carrier,rudno2018control}
For the simulation of carrier and laser dynamics in TI laser devices, the tunneling process is often described via rate 
equations,\cite{bhattacharya1996tunneling,asryan2001tunneling,han2008tunneling,gready2010carrier,gready2011effects} using time constants
extracted from experiments.\cite{bhattacharya2003carrier} Also, the tunneling process itself was the subject of investigations.\cite{chang2004phonon,mielnik2015phonon} 
Based on the calculation of carrier-phonon interaction using perturbation theory, a phonon bottleneck has been predicted, which has led to the 
conclusion that a precise attunement of the level separations of QDs and injector QW to the longitudinal-optical (LO) phonon energy is necessary for tunnel injection devices to operate.

In this paper, we analyze the carrier dynamics of TI-QD laser structures in terms of the interplay between 
structural properties and carrier interaction effects to gauge the engineering possibilities of the TI design and its advantages over conventional laser designs.
Electronic states are determined from three-dimensional {\bf k $\cdot$ p} calculations for the coupled system of injector QW, 
TI barrier, and QDs in order to quantify electronic hybridization effects. 
The influence of QD and QW geometry and material composition on level alignment and hybridization 
strength and the resulting relaxation and capture dynamics of excited carriers due to  carrier-carrier 
Coulomb scattering and carrier scattering with LO-phonons are determined. The carrier dynamics is an important component for efficient laser 
operation, as it controls, e.g., the non-linear gain, the modulation properties, and the temperature stability of the laser.
We analyze the conditions for the level alignment to achieve optimized carrier injection rates.

It is shown that a phonon resonance condition does not appear, for two distinct reasons. 
The electronic states of the joint system of injector QW, TI barrier, and QDs are hybrid states containing a natural spread of energies due to the quasi-continuous nature of the QW states. 
The phenomenon is related to virtual bound states discussed in Ref.~\onlinecite{PhysRevB.80.045327}. 
Furthermore, perturbation theory describes carrier-phonon interaction in terms of free-carrier states while a treatment 
beyond perturbation theory reveals new joint eigenstates of the coupled carrier-phonon system. 
These are  known as polaronic states and can be viewed as hybrid states of the carrier-phonon interaction. 
For LO phonons providing the dominant coupling to lattice vibrations, the inclusion of polaronic effects has been shown effectively lift the resonance criterion and thus avoiding a relaxation bottleneck.\cite{Seebeck:05}
Furthermore, laser devices operate at elevated densities of excited carriers, for which Auger-like carrier-carrier Coulomb scattering processes can provide efficient additional relaxation channels.\cite{Bockelmann:92,Uskov:97,Magnusdottir:03,Nielsen:04}
While our results demonstrate the absence of a relaxation bottleneck, we find a dependency of the injection rate on the level 
alignment for other reasons. The electronic hybridization effect has its own (albeit weak) dependency on the placement of the excited 
QD states with respect to the bottom of the injector QW states. Furthermore, the population distribution of excited carriers in the injector QW 
favors a tuning of the excited QD states to the bottom of the injector QW. With such a design, faster scattering rates into the laser levels 
are obtained in comparison to dot-in-a-well and QD-on-wetting-layer structures. 
TI devices offer a design opportunity to modify the ratio between 
the LO phonon and Coulomb contributions to the carrier dynamics. Our results provide 
tunneling rates as a function of various device parameters to support device modeling.

\section{Electronic structure of the coupled quantum-dot quantum-well system\label{elstcalc}}


\begin{table*}[!ht]
  \centering

  \begin{tabular}{ccc}
    
\begin{tabular}{c}
  \bf (a) \\ \\
  
\begin{tabular}{c|c}
  \bf Region & \bf Material composition \\
  \cline{1-2}
  \\
  Bulk & \\
  InP lattice matched    &   In$_{0.525}$Ga$_{0.235}$Al$_{0.24}$As \\
  \\
  \cline{1-2}
  \\
  Injector QW                    &  In$_{0.695}$Ga$_{0.1525}$Al$_{0.1525}$As 
  \\ \\
  \cline{1-2}
  \\
  Barrier & \\
  InP lattice matched   &   \\ \\
  h = 0 meV  & In$_{0.525}$Ga$_{0.235}$Al$_{0.24}$As \\ \\
  h $\approx$ 175 meV & In$_{0.525}$Ga$_{0.165}$Al$_{0.31}$As \\
  \\
  \cline{1-2}
  \\
  WL                            &   In$_{0.8}$Ga$_{0.1}$Al$_{0.1}$As  \\
  \\
  QD                            &   In$_{0.9}$Ga$_{0.05}$Al$_{0.05}$As \\
  \\
\end{tabular}

\end{tabular}

 & ~~~ &

\begin{tabular}{c}

  \bf (b) \\ \\
  
\begin{tabular}{c|c}
  \bf Injector QW material composition x & \bf Injector QW-QD energy gap \\
  for In$_{x}$Ga$_{(1-x)/2}$Al$_{(1-x)/2}$As & (QD geometry: 16 x 24 nm) \\ \\
  \cline{1-2}
  \\
0.705 & 29 meV \\
0.695 & 34 meV \\
0.689 & 39 meV \\
0.682 & 44 meV \\
\end{tabular}

\\ \\ \\ \bf (c) \\ \\

\begin{tabular}{c|c}
  \bf QD geometry & \bf Injector QW  material composition x\\
              & for In$_{x}$Ga$_{(1-x)/2}$Al$_{(1-x)/2}$As \\ \\
  \cline{1-2}
  \\
16 x 16 nm & 0.675 \\   
16 x 24 nm & 0.695 \\
16 x 32 nm & 0.705 \\
\end{tabular}

\\ \\ \\

\end{tabular}

\end{tabular}

  \caption{(a) Material composition of the TI structure regions depicted in Figs.~\ref{fig1} and \ref{fig2}. (b) Material composition of different injector QW-QD energy gaps $\Delta E^{\text{QW}-\text{QD}}$ as discussed in Fig.~\ref{fig4}, and (c) of the injector QW for different QD geometries illustrated in Fig.~\ref{test1}.}
\label{matcomp1}
\end{table*}


\begin{figure}[tb]
\centering
\includegraphics[trim=7cm 1cm 7cm 2cm,clip,scale=0.7,angle=0]{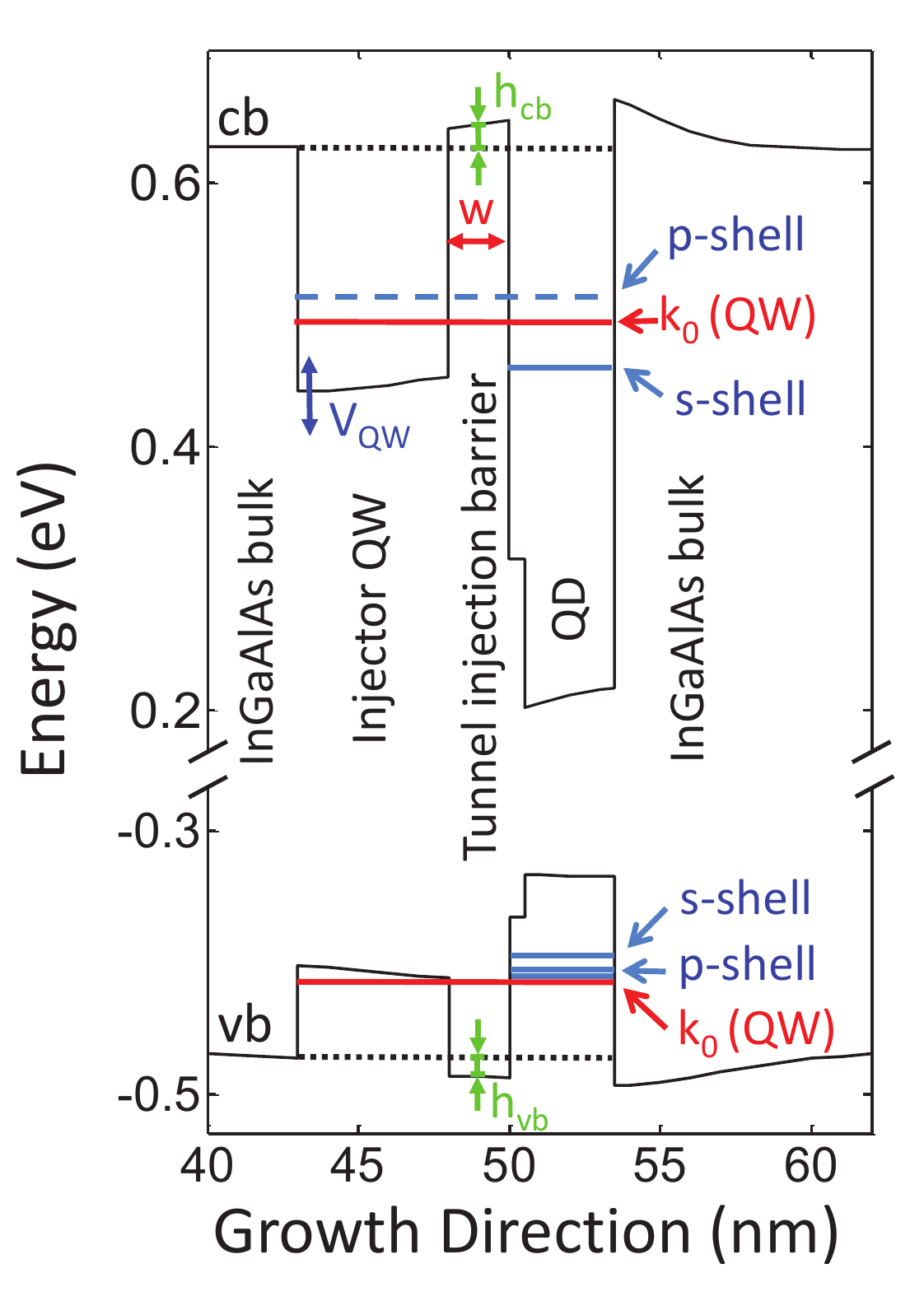}
\caption{Confinement potential of the tunnel-injection structure in the growth direction. The injector QW is separated by a barrier of width $w=2$~nm from the QD. The energy levels of the QD as well as the bottom of the QW-like continuum states $k_{0}$ are presented by horizontal lines. The QD ground-state has an energy difference of 34~meV to the bottom of the QW-like continuum states.}
\label{fig1}
\end{figure}

\begin{figure}[b!]
\includegraphics[trim=2.2cm 1cm 8cm 1cm,clip,scale=0.61,angle=0]{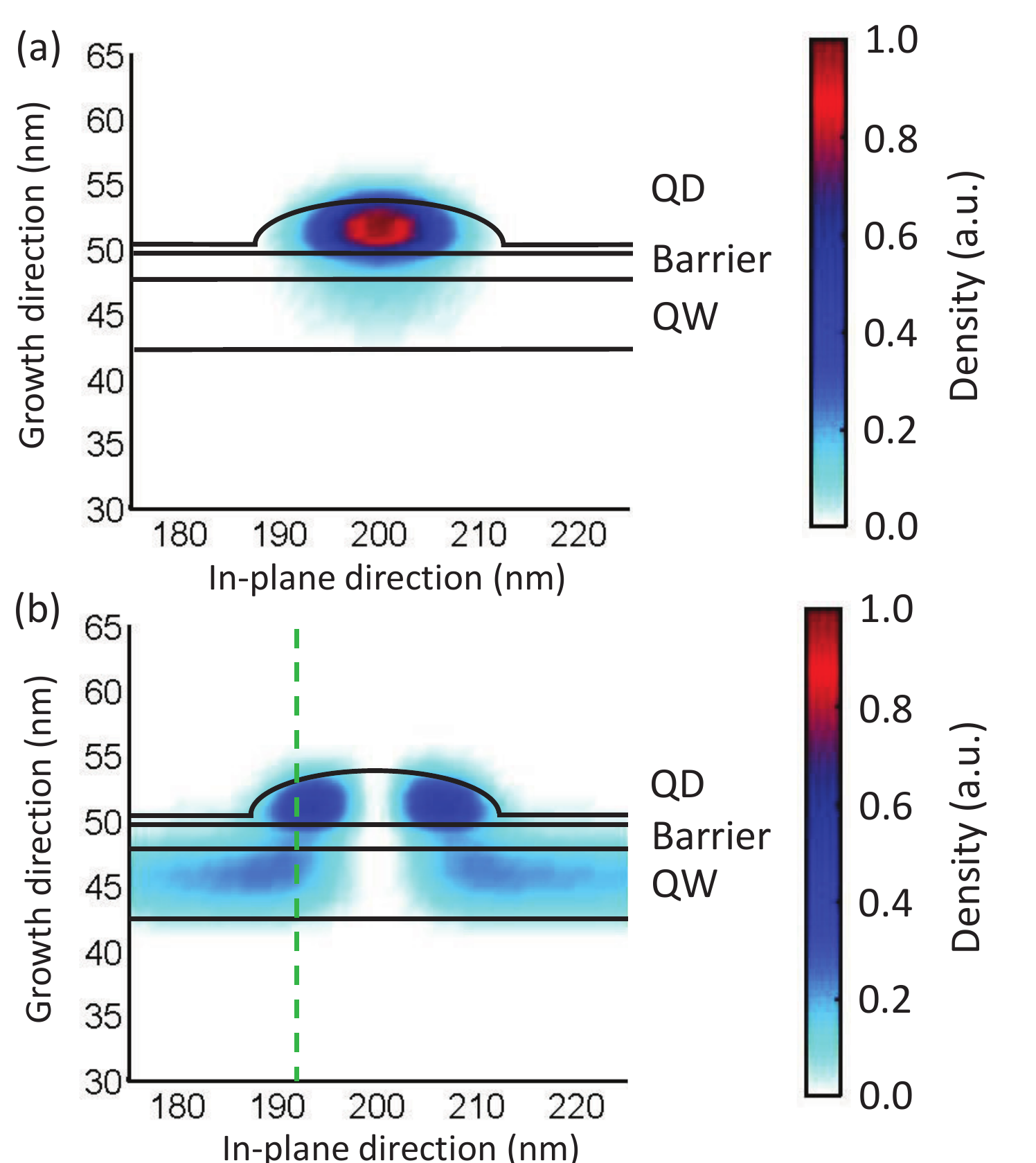}
\caption{Profile of the probability distribution for (a) the QD electron ground state and (b) a hybridized QW-like continuum state for a barrier width of 2~nm.}
\label{fig2}
\end{figure}

\begin{figure}[htb]
\centering
\includegraphics[trim=7cm 0cm 7cm 1cm,clip,scale=0.6,angle=0]{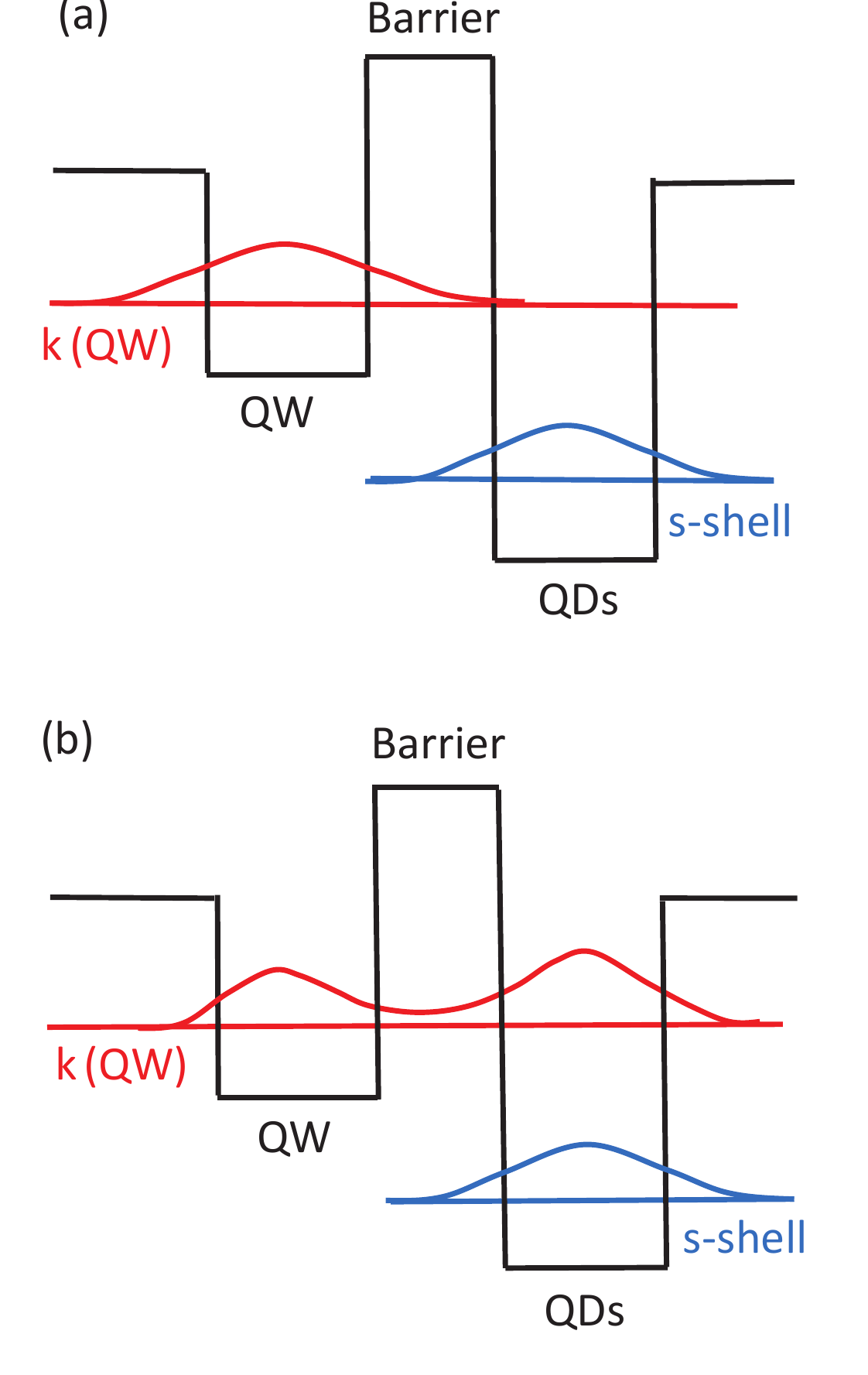}
\caption{(a) Schematic wave functions for QD ground state and a QW state, showing only leaking of the wave functions into the barrier region. 
(b) For hybridized states, the overlap between QD and QW state is much larger than in (a).}
\label{fig0}
\end{figure}

For TI laser structures, 
the injector QW is placed in the growth direction below a QD and separated by a tunnel barrier, as sketched in Fig.~\ref{fig1}. 
To investigate the influence of design properties on the carrier injection rate into the QD ground state, the morphology of the system is varied as follows.
For the injector QW, different material compositions are used to modify its confinement potential depth and the resulting two-dimensional (2D) band edge. 
The tunnel barrier itself can be varied in its width and, via its composition, in its potential height. This  modifies the electronic coupling of QW and QD states
and determines the strength of the hybridization (see below).
Lastly, variations of the QD size can be used for controlling its energy levels, which includes level spacing among the QD states and relative to the injector QW states. 
With the nextnano$^3$ package,\cite{nextnano3} first the spatial strain field distribution is determined, 
which serves as input for the subsequent electronic state calculation via the Pikus-Bir Hamiltonian in the {\bf k  $\cdot$ p}-method.
Furthermore, periodic boundary conditions with large in-plane periodicity are necessary to treat the continuum of the QW-like states.

For a material-realistic description of an existing system, we consider InAs QDs with quarternary barriers lattice matched 
to an InP substrate. A structural model of such a system, which includes QD geometry as well as the alloy concentrations 
for QDs and barrier regions and their spatial variation inside and around the QD, is deduced from recent high-resolution scanning transmission electron microscopy (HRSTEM) 
measurements in Ref.~\onlinecite{rudno2018control}. The used material compositions are summarized in Table \ref{matcomp1}(a). 
We consider a 5~nm width of injector QW and ellipsoidal QDs with 16~nm minor axis, 24~nm major axis, and 3~nm height (unless otherwise 
noted in the text) on a thin monolayer wetting layer (WL). 
The effective confinement potential including strain is shown in Fig.~\ref{fig1}.
The results of the strain can be seen as band bending, e.g., on the right side of
the QD potential.
The small edge in the confinement on the left side of the QD denotes the wetting layer (WL).
For completeness, we note that the WL is too thin to hold confined states.
Additional information regarding the structure can be found in Appendix~\ref{appA}.

For the considered system, the QDs contain one confined state for electrons and three confined states for holes. 
The QD ground state level (s shell in Fig.~\ref{fig1})
is positioned about one LO-phonon energy below the band edge of the injector QW states (marked by $k_0$ in Fig.~\ref{fig1}).
For this design, the electron p state of the QD hybridizes with the energetically nearby states of the QW continuum and the resulting wave functions are partially localized in both QW and QD. 
This is shown in Fig.~\ref{fig2}, where a cut through the wave function reveals the nodal structure of a p state and
also a strong contribution in the injector QW. As we find no node in the growth direction (along the dashed {green} line), we can infer that this is a
bonding rather than an antibonding state.
This coupling mechanism is completely analogous to QD molecules or coupled potential wells like in quantum cascade structures. Figure~\ref{fig2}(a) also reveals that 
the QD s state leaks into the QW region, reemphasizing the hybridized nature of the states. 
For the valence band, the qualitative analysis is similar, but the energy spacing of the states is reduced (due to a higher effective mass). 
Therefore, three discrete QD states are confined. Also the variety of available QD states in the QW-QD continuum affects the hybridization, 
e.g., by providing a higher diversity of hybridized states. It should be noted that not all resulting states of the QW continuum are hybridized states.
Schematically, this situation is depicted in Fig.~\ref{fig0}, showing hybridized states [Fig.~\ref{fig0}(b)] as well as states 
where the overlap between QD and QW states (and the resulting tunneling)
is only due to leaking of the wave functions into the tunnel barrier region [Fig.~\ref{fig0}(a)]. 
In the first case, the overlap of the QD and QW wave function is much higher, leading to more efficient carrier injection.
If phenomenological wave-function models are used to describe the tunneling 
process,\cite{chang2004phonon} often only the leaking wave functions [Fig.~\ref{fig0}(a)] are included.

The foregoing discussion leads to the following picture of the tunnel injection process.
Carriers are captured from bulk states into the injector QW and thermalize via Coulomb and carrier-phonon interaction. This populates the
hybridized states, which are partially localized within the QD. As a final step, from these states, carrier scattering into the QD ground state occurs.

\section{Theory of carrier dynamics in tunnel injection structures\label{theocarrdyn}}
\subsection{Perturbative and non-perturbative treatment of carrier-phonon scattering\label{pertur}}
Using perturbation theory and scattering rates based on Fermi's golden rule, kinetic equations for the
dynamics of the carrier occupation probabilities $f_\alpha$ of the electronic states $\alpha$
can be derived, which contains Boltzmann scattering rates,\cite{CALLAWAY:1991611}
\begin{equation}\label{eq:scattering2}
\begin{split}
  \frac{d}{dt}f_\alpha= \sum\limits_\beta f_\beta  (1-f_\alpha) |M_{\beta\alpha}|^2 \Big((1+n_{\text{LO}}) \delta(\epsilon_{\beta}-\epsilon_{\alpha}-\hbar\omega_{\text{LO}})\\
+ n_\text{LO}\delta(\epsilon_\beta-\epsilon_\alpha+\hbar\omega_{\text{LO}})\Big)\\
					 - f_\alpha (1-f_\beta)  |M_{\alpha\beta}|^2 \Big((1+n_{\text{LO}}) \delta(\epsilon_{\alpha}-\epsilon_{\beta}-\hbar\omega_{\text{LO}})\\
+ n_\text{LO}\delta(\epsilon_\alpha-\epsilon_\beta+\hbar\omega_{\text{LO}})\Big)~.
\end{split}
\end{equation}
Here, $M_{\beta\alpha}$ are the matrix elements of the carrier-phonon interaction, $n_\text{LO}$ is the phonon occupation, $\epsilon_{\alpha}$ is the energy of the state $\alpha$,
and $\omega_{\text{LO}}$ is the phonon frequency. The $\delta$-function in Eq.~\eqref{eq:scattering2} ensures exact energy conservation in the scattering process. 
The kinetic equation contains all possible in- and out-scattering processes between states $\alpha$ and $\beta$ due to LO-phonon emission and absorption processes. 
When applied to the tunnel-injection system, this would require a tuning of the considered electronic levels (QD ground state and the excited QD state, 
which is coupled to the injector QW, see Fig. 3) to the LO-phonon energy to achieve efficient carrier scattering.  The resonance condition is known as the phonon bottleneck and is
the basis of several earlier investigations.\cite{bhattacharya1996tunneling,asryan2001tunneling,han2008tunneling,gready2010carrier} 

A treatment of the carrier-phonon interaction beyond perturbation theory has been introduced in the past in Refs.~\onlinecite{Inoshita:97,Kral:98}  using the 
nonequilibrium Green's functions technique. From a Dyson equation, a generalization of Eq.~\eqref{eq:scattering2} can 
be derived,\cite{Haug_Koch:04} which represents a quantum kinetic equation, 
\begin{equation}\label{eq:qk-scat}
\begin{split}
\frac{\partial f_{\alpha}(t)}{\partial t} = 2\text{ Re}\sum_{\beta}\int\limits^t_{-\infty}dt'~|M_{\alpha\beta}|^2
 &G^{\ret}_{\beta}(t-t')\left[G_{\alpha}^{\ret}(t-t')\right]^* \\
 \ast\, \Big\{\left[f_{\beta}(t')(1-f_{\alpha}(t'))\right] i\hbar \Big[&(1+n_\text{LO})e^{- i\omega_\text{LO}(t-t')}\\
& +n_\text{LO}e^{+ i\omega_\text{LO}(t-t')}\Big] \\
 -\left[f_{\alpha}(t')(1-f_{\beta}(t'))\right]  i\hbar \Big[&(1+n_\text{LO})e^{+ i\omega_\text{LO}(t-t')}\\
& +n_\text{LO}e^{- i\omega_\text{LO}(t-t')}\Big]\Big\}~,
\end{split}
\end{equation}
in which two principal modifications are present. The $\delta$-function, which represents strict energy conservation, is replaced by functions that contain the spectral properties of electrons 
and holes under the influence of the interaction. This includes renormalization effects in the form of quasiparticle energies and their broadening. 
Quasiparticles as eigenstates of the interacting system are characterized by new energies in comparison to the free-particle energies. 
The broadening reflects the finite lifetime of the quasi-particle, caused by
emission and reabsorption of phonons.
In the nonperturbative regime, the new energies reflect the dressing of the electronic states with a series of phonon 
replica due to emission and absorption processes. It is the overlap of these dressed states and their quasiparticle broadening, which is lifting the phonon resonance condition.
Furthermore, non-Markovian effects are included via the t' integral and the explicit dependence of the population functions on the system evolution in the past. 
Non-Markovian effects reflect the finite built-up time of quasiparticles and the influence of this built-up process on scattering rates.
In systems with a quasicontinuous electronic density of states, such as bulk semiconductors or QWs, the perturbative treatment of carrier-phonon interaction is often 
a good approximation, while the nonperturbative treatment provides quantitative corrections 
to the scattering rates - in particular to the ultrafast carrier dynamics.\cite{Banyai:98,Gartner:02}

On the other hand, for QDs it has been shown, that a strong-coupling situation can be realized when the bosonic LO phonons interact with discrete electronic states.\cite{Inoshita:97,Kral:98} The situation resembles the Jaynes-Cummings interaction with a monochromatic light field. The carrier-phonon coupling is enhanced by the electronic-state confinement. In these situations,
the nonperturbative treatment of carrier scattering strongly deviates from perturbative results.\cite{Seebeck:05} While perturbation theory predicts a strong dependence of the scattering rate on the transition energies matching the LO-phonon energy, nonperturbative calculations reveal efficient carrier scattering even if the electronic energy difference strongly departs from the LO-phonon energy, as a particular electronic state can hybridize with a state spaced approximately one LO-phonon energies apart. 
This affects both scattering between QD states and carrier capture from QW states into QD states.\cite{Seebeck:05} For further details of the theoretical model, we refer to Appendix \ref{appB}.

\subsection{Resonance condition for LO-phonon scattering}
  
In the following, we discuss the physics behind the absence of a resonance condition for the LO-phonon-assisted carrier 
scattering from the injector QW and upper QD states into the QD ground state. 
One origin lies in the electronic states of the coupled system. 
Instead of calculating just the overlap of unperturbed quantum-well and quantum-dot states through the tunnel barrier, 
we determine new eigenstates of the coupled system. For the considered level alignment, the excited quantum-dot 
state hybridizes with the quasi-continuum of quantum-well states. 
As a result, the discrete upper quantum-dot state is replaced by a band of electronic energies with a width of several meV. 
This itself lifts the resonance condition for scattering from the hybrid injector quantum well and 
excited quantum-dot state into the discrete quantum-dot ground state due to LO phonons.

As a second origin, Fermi's golden rule (which has led to the prediction of a phonon bottleneck) should not be applied to the 
coupling of discrete quantum-dot states by LO phonons (see, e.g., Seebeck et al. \onlinecite{Seebeck:05}). 
A nonperturbative treatment of the carrier-phonon interaction provides quasiparticle (polaron) 
effects that broaden the individual levels, thereby weakening the resonance condition.

The combination of electronic-state hybridization and nonperturbative carrier-phonon interaction essentially negates 
the resonance criterion to a very large degree, as can be seen from the results in the following section; see especially 
Figs.~\ref{fig4}(a) and~\ref{fig4}(f). Also at high excited carrier densities relevant for laser action, Coulomb scattering complements the carrier-phonon scattering. 
As Coulomb scattering has no preferred energy resonance, this is just another contribution known to lift the LO-phonon bottleneck.

\subsection{Carrier-carrier Coulomb scattering\label{secCOUL}}
In analogy to the carrier-phonon interaction, a kinetic equation for Coulomb scattering can be formulated as 
\begin{multline}\label{eq:qk-scat-coul}
\frac{\partial f_{\alpha}(t)}{\partial t} = 
2\hbar^{2} \text{ Re}\sum_{\beta\gamma\delta}\int\limits^t_{-\infty}dt'
\left|W_{\alpha\gamma\delta\beta}\right|^2 \\
G^{\ret}_{\beta}(t-t')\left[G_{\alpha}^{\ret}(t-t')\right]^* G^{\ret}_{\delta}(t-t')\left[G_{\gamma}^{\ret}(t-t')\right]^*\\
\ast\Big[ f_\beta(t')  (1-f_\alpha(t')) f_\delta(t') (1-f_\gamma(t'))\\
- f_\alpha(t')  (1-f_\beta(t')) f_\gamma(t') (1-f_\delta(t')) \Big]~.
\end{multline}
For the Coulomb scattering, we apply a quasiparticle approximation for the retarded Green's functions (GFs) using polaronic energies and the corresponding quasiparticle broadening. 

Within the Markov approximation, Eqs.~\eqref{eq:qk-scat} and \eqref{eq:qk-scat-coul} can be written in the form of Boltzmann scattering rates
\begin{equation}\label{eq:scattering1}
  \frac{d}{dt}f_\alpha=(1-f_\alpha)\Gamma^{in}_\alpha -f_\alpha\Gamma^{out}_\alpha~,
\end{equation}
where $\Gamma^{in}_\alpha$ and $\Gamma^{out}_\alpha$ describe the in and out scattering of the state $\alpha$. 
For a small deviation from quasiequilibrium, the scattering time $\tau$ can be calculated according to 
\begin{equation}\label{eq:sctime}
  \tau=\left( \Gamma^{in}_\alpha +\Gamma^{out}_\alpha \right)^{-1}~.
\end{equation}

\section{Carrier dynamics \label{carrdyn}}

\subsection{Phonon-mediated tunneling \label{ThElPh}}

In the following, we describe our numerical results for the carrier scattering. To identify their relative importance, 
the carrier-phonon and the Coulomb interaction are investigated separately in the next two sections.
Scattering processes are described in terms of hybridized 
QD-QW states, as discussed in Sec.~\ref{elstcalc}. We utilize the nonperturbative treatment of the carrier-phonon interaction via Eq.~\eqref{eq:qk-scat},
including quasiparticle and non-Markovian effects that reflect the additional hybridization of the electronic states with the phonons.

\begin{figure*}[!htb]
\centering
\includegraphics[trim=2.5cm 4.75cm 3cm 7cm,clip,scale=0.38,angle=0]{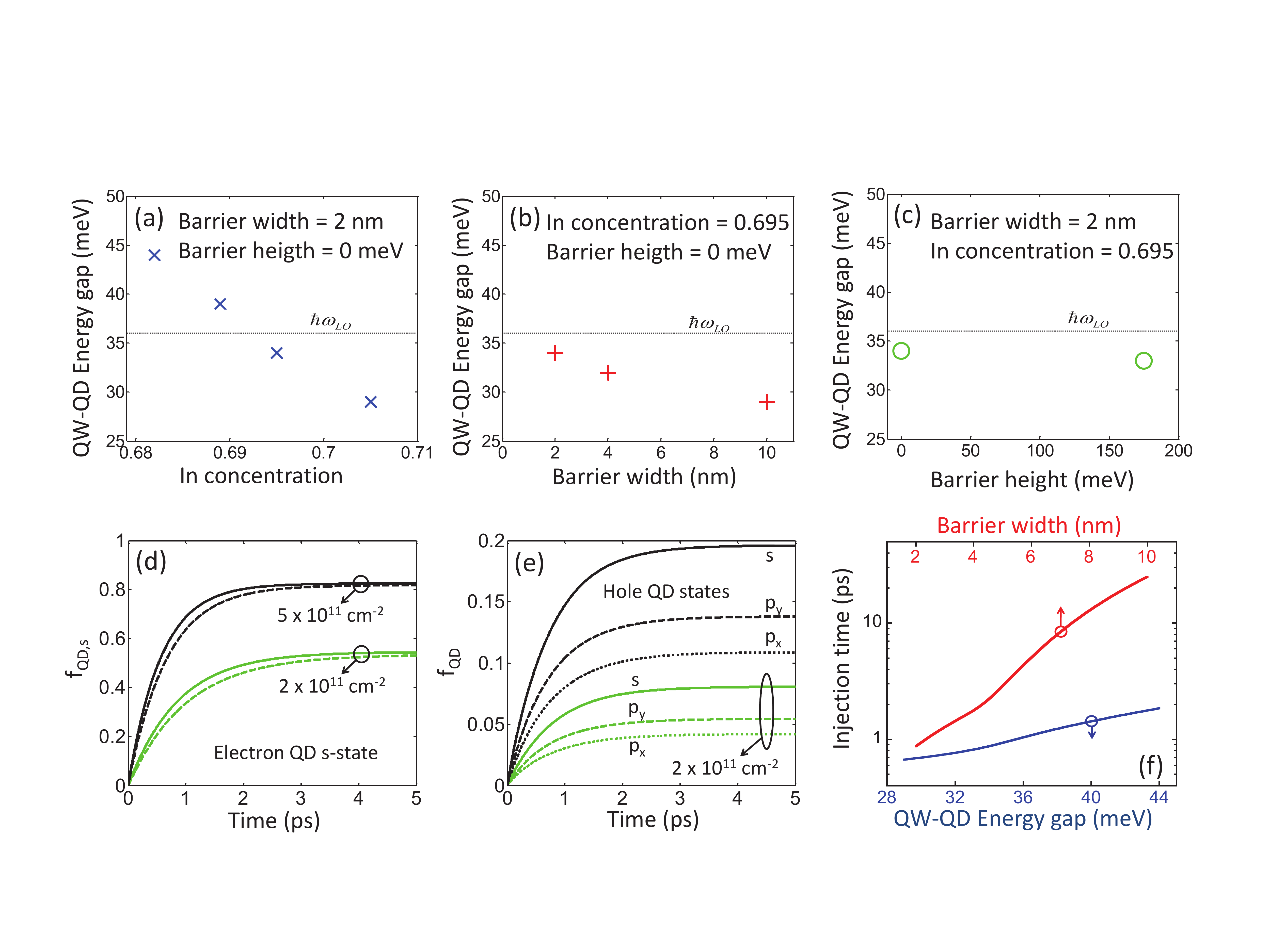}
\caption{Energy difference between the QD ground state and the bottom of the QW-like continuum states vs~(a) material composition of the QW, (b) barrier width, and (c) barrier height.
(d) Population of the QD electron ground state vs time for
phonon-mediated QW to QD relaxation using 2~nm
barrier width and an excess barrier height h$_{cb}$+h$_{vb}$
(relative to the InGaAlAs bulk material; see Fig.~\ref{fig1}) of 0~meV
(solid lines) and 175~meV (dashed lines). (e) Population dynamics
of the QD hole states for 2 nm barrier width
and 175~meV excess barrier height. In (d) and (e)
carrier densities of $2\times 10^{11}\text{cm}^{-2}$ (green lines) and $5\times 10^{11}\text{cm}^{-2}$ (black lines)
are compared.
(f) Carrier injection times for varying barrier width and QW material composition [see (a) and (b)].
\label{fig4}}
\end{figure*}

The TI scheme provides multiple possibilities to tune the system. 
In Figs.~\ref{fig4}(a)-(c), we depict the variations that will be investigated in the following. 
Here, each data point corresponds to a structural design for which
the electronic states, as explained in Sec.~\ref{elstcalc}, and the carrier dynamics, as shown in Sec.~\ref{theocarrdyn}, are evaluated. 
The injector QW can be varied in its composition (blue crosses), modifying the energetic distance between QW and QD states. 
The tunnel barrier can be modified in its width (red crosses) and height (green circles). This controls the strength of the hybridization.
The $y$-axis in Figs.~\ref{fig4}(a)-(c) gives the resulting energetic difference between the QW and 
QD ground states, $\Delta E^{\text{QW}}_{\text{QD}}=E^e_{\bf{k}=0}-E^e_s$.
As a figure of merit, we investigate the capture from electrons
from the injector QW into the QD, assuming a 300~K quasiequilibrium
distribution in the injector QW as well as an
initially empty QD ground state.

The temporal evolution of the electron population for the QD
ground state and the hole population for the three lowest QD states
are shown in Figs.~\ref{fig4}(d) and \ref{fig4}(e), respectively.  Two different carrier
densities below (green lines: $2 \times 10^{11}$ cm$^{-2}$) and at the onset of
optical gain (black lines: $5 \times 10^{11}$ cm$^{-2}$) have been used. We find no
significant variation of the scattering time with carrier density,
as expected for the carrier-phonon interaction. The results in Fig.~\ref{fig4}(d)
demonstrate that for a barrier width of 2~nm and excess
barrier heights (relative to the surrounding bulk material)
of 0~meV (solid lines) and 175~meV
(dashed lines), efficient carrier injection into the QD
state is possible due to strong hybridization effects, even
though in neither case is a matching between the QW-QD
energy and the LO-phonon energy present. The height variation
has only a minor influence on the injection efficiency.
The hole scattering is slightly faster than the electron
scattering, especially due to the decreased level spacing
between the QD and QW-like continuum states.  In Fig.~\ref{fig4}(f), the injection 
time extracted with the help of an exponential fit function for varying barrier width (red) and QW material composition (blue) is shown. 
The barrier width, directly controlling the hybridization efficiency, strongly influences the injection time, in agreement with recent experimental results in Refs.~\onlinecite{bauer2018growth,rudno2018control}.
The QW-QD energy gaps [corresponding to the blue crosses in Fig.~\ref{fig4}(a)] range from significantly below to above the LO phonon resonance [horizontal line in Fig.~\ref{fig4}(a)]. 
As long as the level spacing is below the LO phonon energy, the phonon-mediated capture is very efficient due to strong hybridized QW-QD states.
For energy spacings above the LO phonon resonance,
the efficiency of the capture process decreases as
hybridization effects are reduced and scattering
mediated by polaronic effect is also diminishing.
However, for a detuning of about 10-20~meV from the LO phonon energy, the injection process is still efficient.

The results in Fig.~\ref{fig4} show that for intermediate changes of the tunnel barrier and injector QW geometry and composition, the
influences on the phonon-mediated tunneling strength are small and not strongly dependent on the 
energetic alignment to the phonon energy as often assumed in the literature.\cite{bhattacharya2003carrier,fathpour2005high,bhattacharya2002tunnel,gready2010carrier}

\begin{figure}[t!]
\includegraphics[trim=3.5cm 2cm 2cm 3cm,clip,scale=0.5,angle=0]{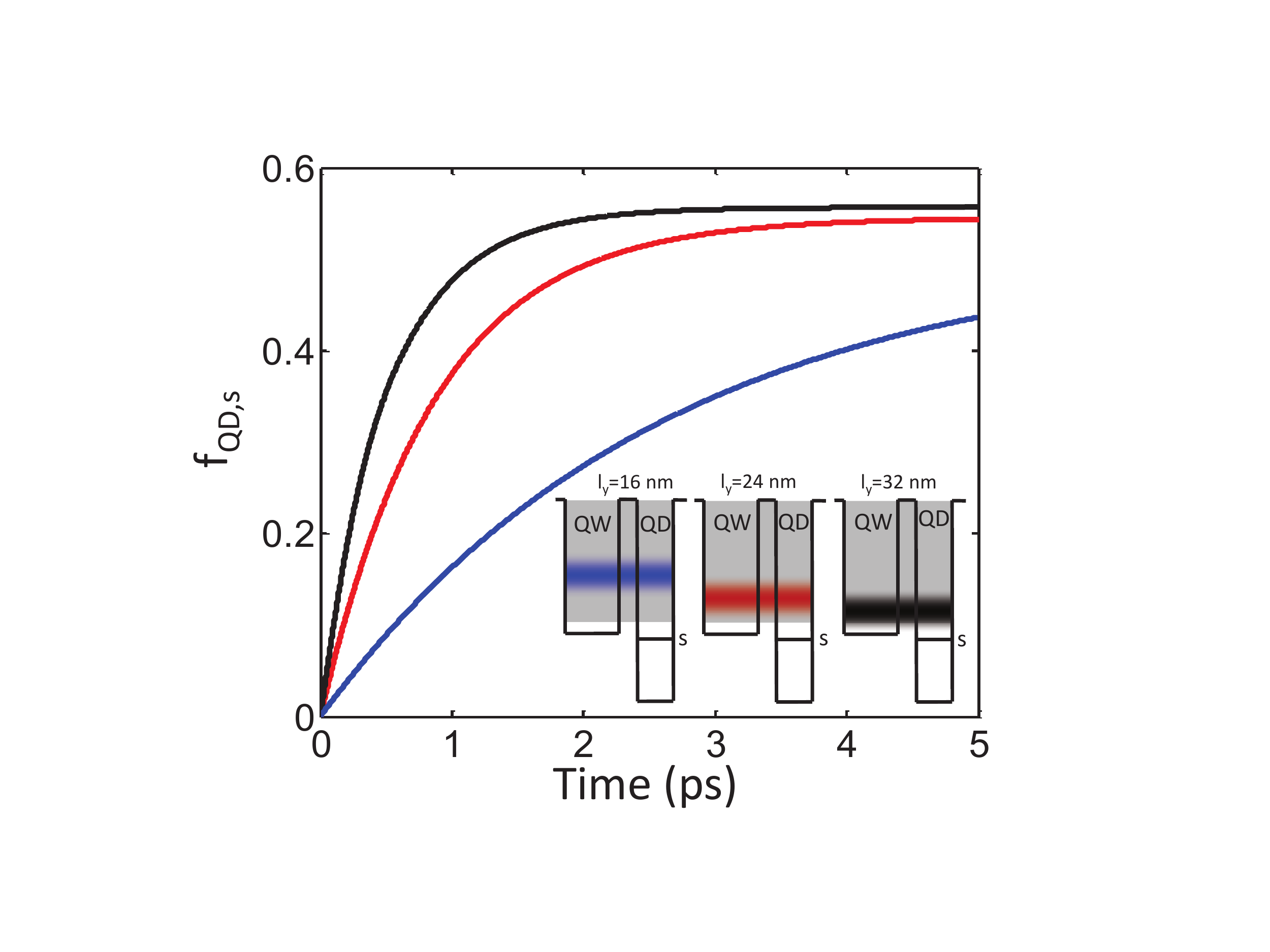}
\caption{Population dynamics of the QD electron ground state vs time for the phonon-mediated relaxation of electrons from the QW into the QD ground state for different QD sizes, leading to different hybridization scenarios.}
\label{test1}
\end{figure}

\begin{figure}[t!]
\includegraphics[width=0.8\columnwidth]{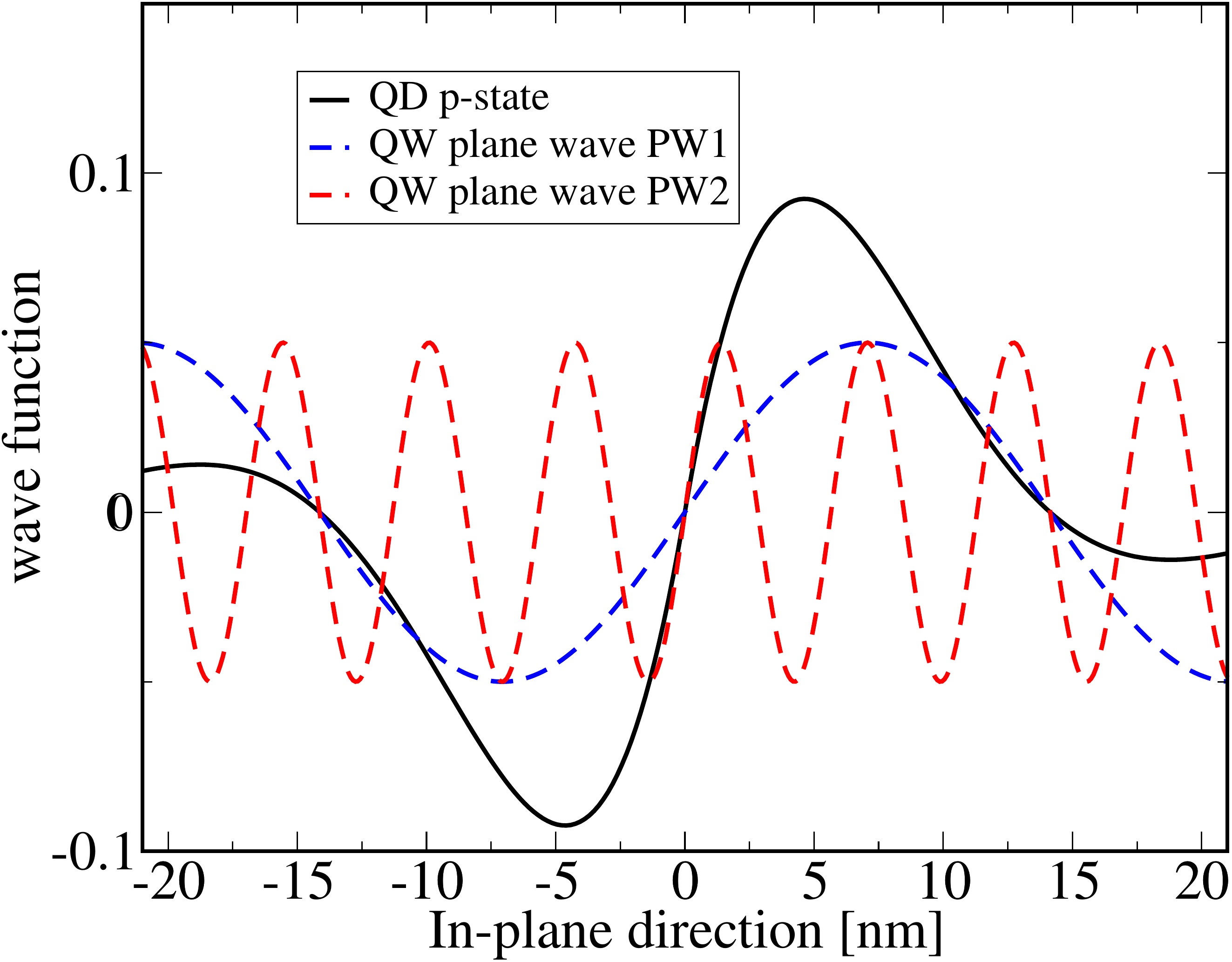}
\caption{Schematic wave function of the QD p state and two different QW plane-wave (PW) states. PW1 has a periodicity that fits well to the QD excited state, 
while PW2 has a much smaller oscillation period. In consequence the hybridization between PW1 and the QD state is more efficient than between PW2 and the QD state. }
\label{fig5}
\end{figure}

While the scattering is not modified significantly by the tuning to the phonon resonance, it does 
depend on the strength of the hybridization. This opens a possibility to tune the injection efficiency, e.g., by modifying the QD geometry.
Such a situation is demonstrated in Fig.~\ref{test1}, where for different QD sizes (varying the long half axis of the ellipsoidal geometry to 16, 24 and 32~nm for fixed short half axis $l_{x}=16$~nm), the phonon mediated tunneling dynamics is shown. The inset depicts the QD ground state, QW conduction band (gray shaded area), and region where hybridization occurs (color shaded area). This comparison reveals that the hybridization is strongly affected by the energetic positions of the QD excited states which are lying energetically in the continuum.
More precisely, for increasing QD size, the energetic spacing between the QD ground and excited state decreases causing the energetic region of hybridized states to move towards the bottom of the conduction band. For the 32 nm QD, it is found at the bottom of the QW conduction band.
To achieve fast carrier scattering, i.e., efficient phonon-mediated tunneling, we see from the results in Fig.~\ref{test1} that the last case is most advantageous. 
This can be understood due to three reasons: First, the population of the injector QW states is highest at the conduction band minimum 
due to the fast intra-QW relaxation. Furthermore, the interaction strength decreases with increasing energy difference between the states 
involved due to the Coulombic nature of the carrier-LO-phonon interaction. In other words, the interaction matrix elements 
are larger if the hybridized part of the conduction band is near the QW band edge.
Finally, the overlap integral between QD and QW states is higher for QW states with lower energy and spatial frequency.
This is schematically shown in Fig.~\ref{fig5} for model wave functions of a lower-energy (PW1) and higher-energy (PW2) QW state and 
a p-like excited QD state. 
In this example, PW1 has a periodicity that leads to a much better overlap integral with the QD p state than PW2 and
hence to a more efficient hybridization.

As a result, the scattering rates of carriers in TI structures from the injector QW into the QDs exhibits a tunability; however, 
it is not governed by the energetic alignment with the phonon energy, but due to the alignment of the excited QD state with the QW 
conduction band minimum that determines the strength of the hybridization.

\subsection{Coulomb contribution to tunneling \label{ThElEl} and scattering times}
\begin{figure}[tb]
\centering
\includegraphics[trim=3cm 3cm 1cm 1cm,clip,scale=0.65,angle=0]{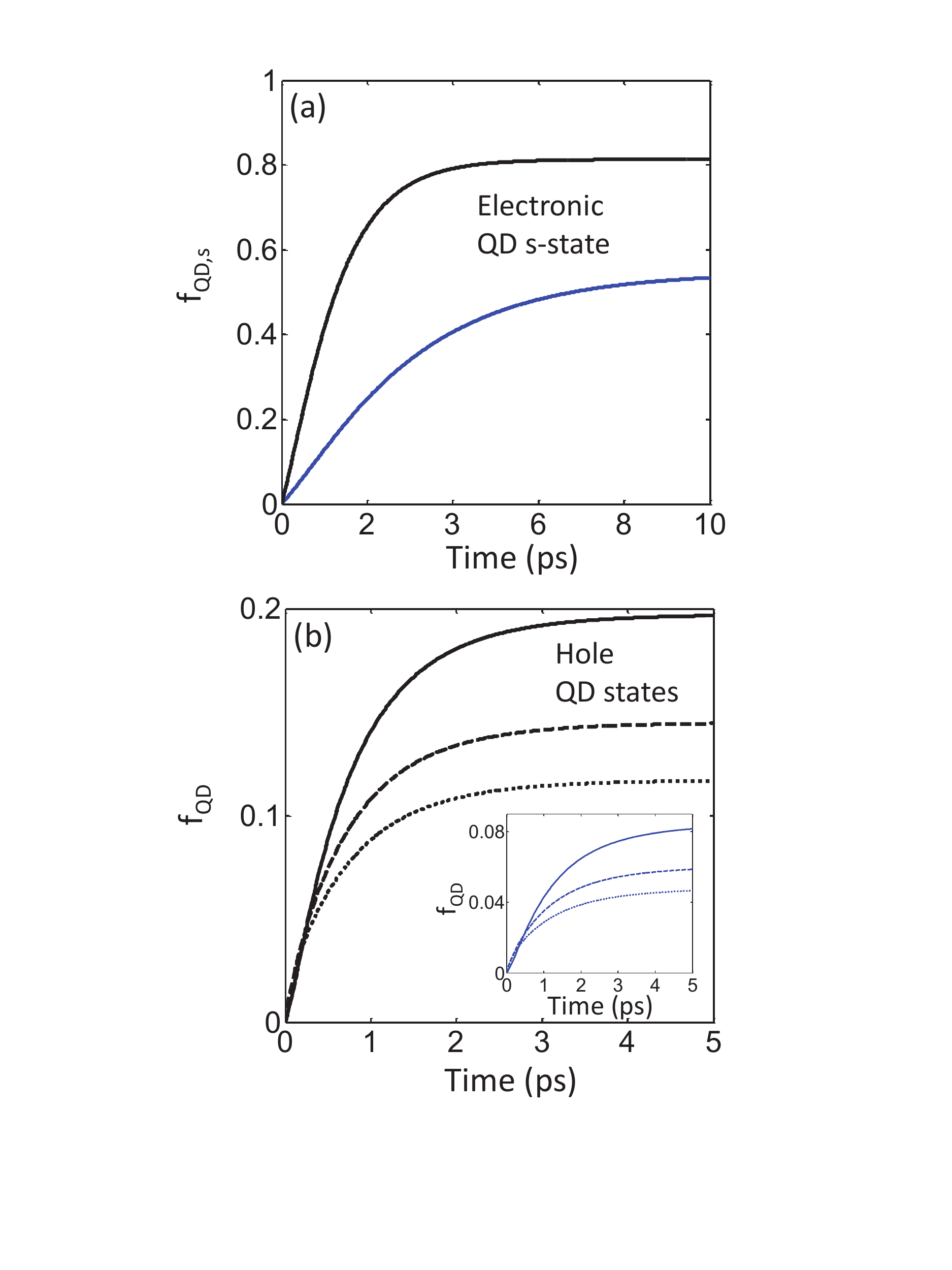}
\caption{Population dynamics of the (a) electron and (b) hole QD ground states (solid line) and excited states (dashed and dotted lines) 
vs time for the Coulomb relaxation process from the QW into the QD for carrier densities of 2 $\times 10^{11}$ cm$^{-2}$ (blue lines or inset) 
and 5 $\times 10^{11}$ cm$^{-2}$ (black lines).}
\label{fig11}
\end{figure}

\begin{figure}[tb]
\centering
\includegraphics[trim=7cm 0.75cm 6cm 0.5cm,clip,scale=0.7,angle=0]{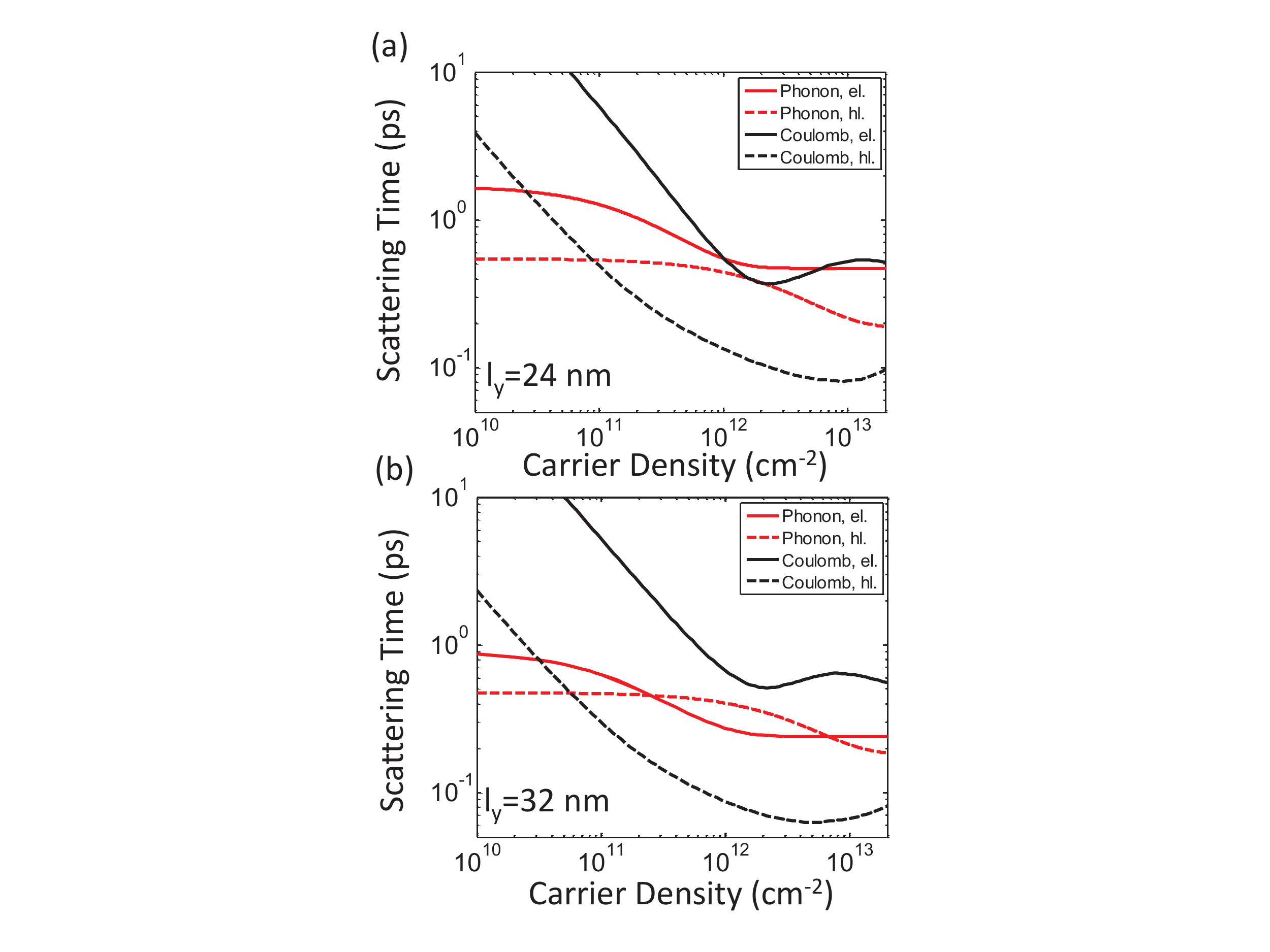}
\caption{Coulomb (black) and phonon-mediated (red) scattering times vs carrier density for the QD electron (solid lines) and hole (dashed lines) states. The QD size is varied by the long half-axis of the ellipsoidal geometry from (a) 24 nm to (b) 32 nm.}
\label{fig13}
\end{figure}

In this section, we investigate the Coulomb scattering contribution to the carrier injection processes. 
As for the phonons, this interaction mechanism is described as scattering between hybridized QD-QW states, 
using a generalized Boltzmann equation including quasiparticle properties as discussed in Sec.~\ref{secCOUL}.
For the Coulomb interaction no significant influence of the detuning is expected due to the absence 
of characteristic energies like the one for LO phonons. We choose a barrier width of 2 nm, an excess barrier height of 175 meV, and an indium concentration of 0.695 
as an example to discuss the influence of the Coulomb mediated tunneling in TI structures. 
To have comparable conditions to Sec.~\ref{ThElPh}, we assume quasiequilibrium distributions with varying 
carrier density at a temperature of 300 K in the injector QW and an initially empty QD ground state.

The results for the Coulomb-mediated injection are shown in Fig.~\ref{fig11}. In contrast to findings for conventional QD laser structures,\cite{Seebeck:05, Nielsen:04,Lorke:06c,Lorke:11} 
the Coulomb-mediated scattering of electrons [Fig.~\ref{fig11}(a)] is significantly slower than the carrier-phonon interaction for a 
carrier density of 2 $\times 10^{11}$ cm$^{-2}$ and even for an elevated carrier density of 5 $\times 10^{11}$ cm$^{-2}$. 
This is caused by the fact that in the TI structure only continuum-assisted capture processes are possible, 
which are known to be much less efficient in comparison to other Coulomb scattering processes.\cite{Nielsen:04} For holes, 
the multitude of scattering channels causes fast Coulomb scattering for both carrier densities in Figs.~\ref{fig11}(b). 
As in the case of carrier-phonon interaction, the steady-state populations are significantly higher for electrons than for holes due to the higher effective mass of the latter.

To facilitate the use of the presented results in device models based on rate equations, we extract scattering times for the TI structure by considering the relaxation of a small perturbation of a quasi-equilibrium distribution. These scattering times are calculated via Eq.~\eqref{eq:sctime} for two different QD sizes and plotted in Fig.~\ref{fig13} for carrier-phonon and Coulomb interaction. As expected, the Coulomb contribution shows a stronger density dependence than the phonon-mediated coupling.
In agreement with the results of Fig.~\ref{fig11}, the electron-phonon interaction is dominant up to a carrier density of 1 $\times 10^{12}$ cm$^{-2}$ for the 24~nm QD geometry. The reduction of scattering efficiency between carrier densities of 2 $\times 10^{12}$ cm$^{-2}$ - 1 $\times 10^{13}$ cm$^{-2}$ is due to screening of the Coulomb interaction in \eqref{eq:qk-scat-coul}, while for densities above 1 $\times 10^{13}$ cm$^{-2}$, electron-hole Coulomb scattering dominates. Due to the different weighting of the phase space in phonon and Coulomb interaction, the different hybridization scenarios (see Fig.~\ref{test1}) shift the ratio between Coulomb and phonon scattering for the electrons. This is demonstrated by a comparison of the 24 nm with the 32 nm QD geometry in Fig.~\ref{fig13}. In the latter case, the carrier-phonon interaction is dominant for injection of electrons for all densities considered.
For hole scattering the Coulomb interaction becomes more efficient at a carrier density of about  1 $\times 10^{11}$ cm$^{-2}$ due to the efficient hole relaxation and electron-assisted processes. The hybridization of the two QD geometries is more similar for the holes and thus the expected difference between the two QD geometries is less distinct.

\section{Conclusion}
In this paper, the carrier injection from a QW to the QD ground state for TI structures has been investigated. 
The QW-like hybridized states support a fast relaxation into the QD ground state. 
In contrast to previous works, it is found that the tunability of the interaction strength in the TI structure is not governed by the energetic alignment 
with the phonon energy, but by the alignment of the excited QD state with the QW conduction band minimum that determines 
the strength of the hybridization.
Unlike conventional QD laser structures, only continuum-assisted capture processes are possible for the electrons in the TI system. This decreases the importance of the Coulomb interaction relative to the electron-phonon interaction. For structures with strong hybridization the electron-phonon interaction can be the dominant process even at high carrier densities.
The insensitivity of the carrier-phonon interaction to variations of the QW or barrier material composition 
supports the experimental findings of high modulations rates 
due to a reduction of the gain nonlinearity and spectral hole burning.

\begin{acknowledgments}
The authors would like to thank J. P. Reithmaier and G. Eisenstein for fruitful discussions. We acknowledge funding from the DFG and a grant for CPU time from the HLRN (Hannover/Berlin).
\end{acknowledgments}

\appendix 

\section{Electronic structure calculation\label{appA}}

In the TI structure, the injector QW and QDs are separated by a tunnel barrier as depicted in Fig.~\ref{fig1}. 
For a microscopic description of the physics underlying the phonon-mediated tunneling, 
we determine the electronic structure of the coupled injector QW and QD states. A subspace of the continuum states forms hybridized 
injector QW-QD states as discussed in Sec.~\ref{elstcalc}. We find that these hybridized states play a central role when calculating 
the scattering efficiency of carriers from the injector QW into the QDs. Nevertheless, hybridized states are often neglected in phenomenological 
models when considering only the wave function overlap of independent QW and QD states separated by a TI barrier. 

The results in this paper are based on three-dimensional wave-function calculations of the TI structure for the discretized {\bf k $\cdot$ p} Hamiltonian 
including strain and piezoelectric effects; see Ref.~\onlinecite{nextnano3}. To dissolve the continuum states accurately to determine the 
weightage between hybridized and nonhybridized continuum subspaces, we used in-plane dimensions of 400~nm and checked the 
convergence for in-plane dimensions up to 800~nm, thereby calculating over 1000 eigenstates of the coupled QW-QD system. 
Comparing six-band {\bf k $\cdot$ p} to single-band calculations, we find that band mixing plays only a minor role and that the relevant hole states are dominated 
by heavy-hole contributions. As a trade-off between numerical effort and required accuracy, we therefore use a single-band description in practical calculations.

Details of the parameters for the considered structure are provided in Table \ref{matcomp1}.
For the investigations in Fig.~\ref{fig4}, the alloy concentrations of the injector QW are modified according to Table~\ref{matcomp1}.(b) in order to change the level alignment 
and the wave function properties. In Fig.~\ref{test1} variations of the hybridization effect are studied for an 
unchanged energy gap between the injector QW and the QD ground state matching the LO-phonon energy. This is accomplished by combining 
the variation of the QD geometry (to change the hybridization effect) with altered material compositions in the injector QW to keep the 
energy gap unchanged according to Table~\ref{matcomp1}.(c).

\section{Microscopic description of the carrier-phonon interaction\label{appB}}

Single-particle wave functions obtained from electronic structure calculations are used to construct the Coulomb matrix elements according to  \cite{Nielsen:04}
\begin{equation}\label{eq:coulmatr}
V_{\alpha\beta\gamma\delta} = \frac{1}{A} \: \sum \limits_{\bf q} \,
V_{\bf q} \Braket{\alpha| e^{-i {\bf q} \cdot {\bf r}}|\delta}
\Braket{\beta | e^{+i {\bf q} \cdot {\bf r}} |\gamma}
\end{equation}
with the Coulomb potential $V_{\bf q}$. The index $\alpha$ contains states, bands, and spin.
For the screened Coulomb interaction matrix elements $W_{\alpha\beta\gamma\delta}$, we use a generalization of the static Lindhard formula
which is explained in detail in Ref.~\onlinecite{Nielsen:04}. This procedure leads to the replacement
$V_{\bf q}\rightarrow W_{\bf q}$ in Eq.~\eqref{eq:coulmatr}.
The matrix elements for the interaction of carriers with LO phonons are \cite{Seebeck:05,Lorke:06}
\begin{equation}
|M_{\alpha\beta}|^2=\frac{M^2_{LO}}{e^2/\varepsilon_0}V_{\alpha\beta\alpha\beta}~,
\end{equation}
with the prefactor
\begin{equation}\label{eq:prefatorM}
M^{2}_{LO}=4\pi\alpha\frac{\hbar}{\sqrt{2m}}(\hbar\omega_{LO})^{\frac{3}{2}}~
\end{equation}
that includes the polar coupling strength $\alpha$ and the reduced mass $m$.

The functions containing the spectral properties of electrons 
and holes under the influence of the interaction [see Eq.~\eqref{eq:qk-scat}] are the so-called retarded Greens functions.
These follow from a Dyson equation \cite{Seebeck:05} 
\begin{equation}
   \Big[ i\hbar\pdiff - \epsilon_{\alpha} \Big]~G^{\text{\ret}}_{\alpha}(t) =  \delta(t) + \int\!dt' ~~ \Sigma^{\text{\ret}}_{\alpha}(t-t') ~ G^{\text{\ret}}_{\alpha}(t')~.
\label{eq:Dyson_G_ret}
\end{equation}
For the corresponding retarded selfenergy in random-phase approximation (RPA),\cite{Mahan:90} we obtain
\begin{equation}
      \Sigma^{\text{\ret}}_{\alpha}(t) ~ = ~ i \hbar \sum_{\beta} ~|M_{\alpha\beta}|^2
      G^{\text{\ret}}_{\beta}(t) ~ d^<(-t),
\label{eq:Sigma_ret}
\end{equation}
where the phonon propagators $d^\gtrless$ are given by
	\begin{equation}\label{equ:prop}
		d^{\gtrless}(t,~t')=\left[(1+n_\text{LO})e^{\mp i\omega_\text{LO}(t-t')}+n_\text{LO}e^{\pm i\omega_\text{LO}(t-t')}\right]	~,
	\end{equation}
and contain the phonon frequency and the phonon population.
The carrier-phonon scattering is described in terms of a quantum-kinetic equation [see Eq.\eqref{eq:qk-scat}] that includes the retarded GF and contains scattering by phonon-emission and -absorption processes.

\bibliography{mybib2}

\begin{thebibliography}{48}%
\makeatletter
\providecommand \@ifxundefined [1]{%
 \@ifx{#1\undefined}
}%
\providecommand \@ifnum [1]{%
 \ifnum #1\expandafter \@firstoftwo
 \else \expandafter \@secondoftwo
 \fi
}%
\providecommand \@ifx [1]{%
 \ifx #1\expandafter \@firstoftwo
 \else \expandafter \@secondoftwo
 \fi
}%
\providecommand \natexlab [1]{#1}%
\providecommand \enquote  [1]{``#1''}%
\providecommand \bibnamefont  [1]{#1}%
\providecommand \bibfnamefont [1]{#1}%
\providecommand \citenamefont [1]{#1}%
\providecommand \href@noop [0]{\@secondoftwo}%
\providecommand \href [0]{\begingroup \@sanitize@url \@href}%
\providecommand \@href[1]{\@@startlink{#1}\@@href}%
\providecommand \@@href[1]{\endgroup#1\@@endlink}%
\providecommand \@sanitize@url [0]{\catcode `\\12\catcode `\$12\catcode
  `\&12\catcode `\#12\catcode `\^12\catcode `\_12\catcode `\%12\relax}%
\providecommand \@@startlink[1]{}%
\providecommand \@@endlink[0]{}%
\providecommand \url  [0]{\begingroup\@sanitize@url \@url }%
\providecommand \@url [1]{\endgroup\@href {#1}{\urlprefix }}%
\providecommand \urlprefix  [0]{URL }%
\providecommand \Eprint [0]{\href }%
\providecommand \doibase [0]{http://dx.doi.org/}%
\providecommand \selectlanguage [0]{\@gobble}%
\providecommand \bibinfo  [0]{\@secondoftwo}%
\providecommand \bibfield  [0]{\@secondoftwo}%
\providecommand \translation [1]{[#1]}%
\providecommand \BibitemOpen [0]{}%
\providecommand \bibitemStop [0]{}%
\providecommand \bibitemNoStop [0]{.\EOS\space}%
\providecommand \EOS [0]{\spacefactor3000\relax}%
\providecommand \BibitemShut  [1]{\csname bibitem#1\endcsname}%
\let\auto@bib@innerbib\@empty
\bibitem [{\citenamefont {Zhang}\ \emph {et~al.}(1997)\citenamefont {Zhang},
  \citenamefont {Gutierrez-Aitken}, \citenamefont {Klotzkin}, \citenamefont
  {Bhattacharya}, \citenamefont {Caneau},\ and\ \citenamefont
  {Bhat}}]{zhang19970}%
  \BibitemOpen
  \bibfield  {author} {\bibinfo {author} {\bibfnamefont {X.}~\bibnamefont
  {Zhang}}, \bibinfo {author} {\bibfnamefont {A.}~\bibnamefont
  {Gutierrez-Aitken}}, \bibinfo {author} {\bibfnamefont {D.}~\bibnamefont
  {Klotzkin}}, \bibinfo {author} {\bibfnamefont {P.}~\bibnamefont
  {Bhattacharya}}, \bibinfo {author} {\bibfnamefont {C.}~\bibnamefont
  {Caneau}}, \ and\ \bibinfo {author} {\bibfnamefont {R.}~\bibnamefont
  {Bhat}},\ }\href@noop {} {\bibfield  {journal} {\bibinfo  {journal} {IEEE J.
  Sel. Top. Quantum Electron.}\ }\textbf {\bibinfo {volume} {3}},\ \bibinfo
  {pages} {309} (\bibinfo {year} {1997})}\BibitemShut {NoStop}%
\bibitem [{\citenamefont {Bhattacharya}\ \emph {et~al.}(2003)\citenamefont
  {Bhattacharya}, \citenamefont {Ghosh}, \citenamefont {Pradhan}, \citenamefont
  {Singh}, \citenamefont {Wu}, \citenamefont {Urayama}, \citenamefont {Kim},\
  and\ \citenamefont {Norris}}]{bhattacharya2003carrier}%
  \BibitemOpen
  \bibfield  {author} {\bibinfo {author} {\bibfnamefont {P.}~\bibnamefont
  {Bhattacharya}}, \bibinfo {author} {\bibfnamefont {S.}~\bibnamefont {Ghosh}},
  \bibinfo {author} {\bibfnamefont {S.}~\bibnamefont {Pradhan}}, \bibinfo
  {author} {\bibfnamefont {J.}~\bibnamefont {Singh}}, \bibinfo {author}
  {\bibfnamefont {Z.-K.}\ \bibnamefont {Wu}}, \bibinfo {author} {\bibfnamefont
  {J.}~\bibnamefont {Urayama}}, \bibinfo {author} {\bibfnamefont
  {K.}~\bibnamefont {Kim}}, \ and\ \bibinfo {author} {\bibfnamefont {T.~B.}\
  \bibnamefont {Norris}},\ }\href@noop {} {\bibfield  {journal} {\bibinfo
  {journal} {IEEE J. Quantum Electron.}\ }\textbf {\bibinfo {volume} {39}},\
  \bibinfo {pages} {952} (\bibinfo {year} {2003})}\BibitemShut {NoStop}%
\bibitem [{\citenamefont {Fathpour}\ \emph {et~al.}(2005)\citenamefont
  {Fathpour}, \citenamefont {Mi},\ and\ \citenamefont
  {Bhattacharya}}]{fathpour2005high}%
  \BibitemOpen
  \bibfield  {author} {\bibinfo {author} {\bibfnamefont {S.}~\bibnamefont
  {Fathpour}}, \bibinfo {author} {\bibfnamefont {Z.}~\bibnamefont {Mi}}, \ and\
  \bibinfo {author} {\bibfnamefont {P.}~\bibnamefont {Bhattacharya}},\
  }\href@noop {} {\bibfield  {journal} {\bibinfo  {journal} {Journal of Physics
  D: Applied Physics}\ }\textbf {\bibinfo {volume} {38}},\ \bibinfo {pages}
  {2103} (\bibinfo {year} {2005})}\BibitemShut {NoStop}%
\bibitem [{\citenamefont {Bhattacharya}\ and\ \citenamefont
  {Ghosh}(2002)}]{bhattacharya2002tunnel}%
  \BibitemOpen
  \bibfield  {author} {\bibinfo {author} {\bibfnamefont {P.}~\bibnamefont
  {Bhattacharya}}\ and\ \bibinfo {author} {\bibfnamefont {S.}~\bibnamefont
  {Ghosh}},\ }\href@noop {} {\bibfield  {journal} {\bibinfo  {journal} {Appl.
  Phys. Lett.}\ }\textbf {\bibinfo {volume} {80}},\ \bibinfo {pages} {3482}
  (\bibinfo {year} {2002})}\BibitemShut {NoStop}%
\bibitem [{\citenamefont {Mi}\ \emph {et~al.}(2005)\citenamefont {Mi},
  \citenamefont {Bhattacharya},\ and\ \citenamefont {Fathpour}}]{mi2005high}%
  \BibitemOpen
  \bibfield  {author} {\bibinfo {author} {\bibfnamefont {Z.}~\bibnamefont
  {Mi}}, \bibinfo {author} {\bibfnamefont {P.}~\bibnamefont {Bhattacharya}}, \
  and\ \bibinfo {author} {\bibfnamefont {S.}~\bibnamefont {Fathpour}},\
  }\href@noop {} {\bibfield  {journal} {\bibinfo  {journal} {Appl. Phys.
  Lett.}\ }\textbf {\bibinfo {volume} {86}},\ \bibinfo {pages} {153109}
  (\bibinfo {year} {2005})}\BibitemShut {NoStop}%
\bibitem [{\citenamefont {Mi}\ \emph {et~al.}(2006)\citenamefont {Mi},
  \citenamefont {Bhattacharya},\ and\ \citenamefont {Yang}}]{mi2006growth}%
  \BibitemOpen
  \bibfield  {author} {\bibinfo {author} {\bibfnamefont {Z.}~\bibnamefont
  {Mi}}, \bibinfo {author} {\bibfnamefont {P.}~\bibnamefont {Bhattacharya}}, \
  and\ \bibinfo {author} {\bibfnamefont {J.}~\bibnamefont {Yang}},\ }\href@noop
  {} {\bibfield  {journal} {\bibinfo  {journal} {Appl. Phys. Lett.}\ }\textbf
  {\bibinfo {volume} {89}},\ \bibinfo {pages} {153109} (\bibinfo {year}
  {2006})}\BibitemShut {NoStop}%
\bibitem [{\citenamefont {Lee}\ \emph {et~al.}(2011)\citenamefont {Lee},
  \citenamefont {Bhattacharya}, \citenamefont {Frost},\ and\ \citenamefont
  {Guo}}]{lee2011characteristics}%
  \BibitemOpen
  \bibfield  {author} {\bibinfo {author} {\bibfnamefont {C.-S.}\ \bibnamefont
  {Lee}}, \bibinfo {author} {\bibfnamefont {P.}~\bibnamefont {Bhattacharya}},
  \bibinfo {author} {\bibfnamefont {T.}~\bibnamefont {Frost}}, \ and\ \bibinfo
  {author} {\bibfnamefont {W.}~\bibnamefont {Guo}},\ }\href@noop {} {\bibfield
  {journal} {\bibinfo  {journal} {Appl. Phys. Lett.}\ }\textbf {\bibinfo
  {volume} {98}},\ \bibinfo {pages} {011103} (\bibinfo {year}
  {2011})}\BibitemShut {NoStop}%
\bibitem [{\citenamefont {Bhowmick}\ \emph {et~al.}(2014)\citenamefont
  {Bhowmick}, \citenamefont {Baten}, \citenamefont {Frost}, \citenamefont
  {Ooi},\ and\ \citenamefont {Bhattacharya}}]{bhowmick2014high}%
  \BibitemOpen
  \bibfield  {author} {\bibinfo {author} {\bibfnamefont {S.}~\bibnamefont
  {Bhowmick}}, \bibinfo {author} {\bibfnamefont {M.~Z.}\ \bibnamefont {Baten}},
  \bibinfo {author} {\bibfnamefont {T.}~\bibnamefont {Frost}}, \bibinfo
  {author} {\bibfnamefont {B.~S.}\ \bibnamefont {Ooi}}, \ and\ \bibinfo
  {author} {\bibfnamefont {P.}~\bibnamefont {Bhattacharya}},\ }\href@noop {}
  {\bibfield  {journal} {\bibinfo  {journal} {IEEE J. Quantum Electron.}\
  }\textbf {\bibinfo {volume} {50}},\ \bibinfo {pages} {7} (\bibinfo {year}
  {2014})}\BibitemShut {NoStop}%
\bibitem [{\citenamefont {Podemski}\ \emph {et~al.}(2006)\citenamefont
  {Podemski}, \citenamefont {Kudrawiec}, \citenamefont {Misiewicz},
  \citenamefont {Somers}, \citenamefont {Reithmaier},\ and\ \citenamefont
  {Forchel}}]{podemski2006tunnel}%
  \BibitemOpen
  \bibfield  {author} {\bibinfo {author} {\bibfnamefont {P.}~\bibnamefont
  {Podemski}}, \bibinfo {author} {\bibfnamefont {R.}~\bibnamefont {Kudrawiec}},
  \bibinfo {author} {\bibfnamefont {J.}~\bibnamefont {Misiewicz}}, \bibinfo
  {author} {\bibfnamefont {A.}~\bibnamefont {Somers}}, \bibinfo {author}
  {\bibfnamefont {J.~P.}\ \bibnamefont {Reithmaier}}, \ and\ \bibinfo {author}
  {\bibfnamefont {A.}~\bibnamefont {Forchel}},\ }\href@noop {} {\bibfield
  {journal} {\bibinfo  {journal} {Appl. Phys. Lett.}\ }\textbf {\bibinfo
  {volume} {89}},\ \bibinfo {pages} {061902} (\bibinfo {year}
  {2006})}\BibitemShut {NoStop}%
\bibitem [{\citenamefont {S{\k{e}}k}\ \emph {et~al.}(2007)\citenamefont
  {S{\k{e}}k}, \citenamefont {Poloczek}, \citenamefont {Podemski},
  \citenamefont {Kudrawiec}, \citenamefont {Misiewicz}, \citenamefont {Somers},
  \citenamefont {Hein}, \citenamefont {H{\"o}fling},\ and\ \citenamefont
  {Forchel}}]{skek2007experimental}%
  \BibitemOpen
  \bibfield  {author} {\bibinfo {author} {\bibfnamefont {G.}~\bibnamefont
  {S{\k{e}}k}}, \bibinfo {author} {\bibfnamefont {P.}~\bibnamefont {Poloczek}},
  \bibinfo {author} {\bibfnamefont {P.}~\bibnamefont {Podemski}}, \bibinfo
  {author} {\bibfnamefont {R.}~\bibnamefont {Kudrawiec}}, \bibinfo {author}
  {\bibfnamefont {J.}~\bibnamefont {Misiewicz}}, \bibinfo {author}
  {\bibfnamefont {A.}~\bibnamefont {Somers}}, \bibinfo {author} {\bibfnamefont
  {S.}~\bibnamefont {Hein}}, \bibinfo {author} {\bibfnamefont {S.}~\bibnamefont
  {H{\"o}fling}}, \ and\ \bibinfo {author} {\bibfnamefont {A.}~\bibnamefont
  {Forchel}},\ }\href@noop {} {\bibfield  {journal} {\bibinfo  {journal} {Appl.
  Phys. Lett.}\ }\textbf {\bibinfo {volume} {90}},\ \bibinfo {pages} {081915}
  (\bibinfo {year} {2007})}\BibitemShut {NoStop}%
\bibitem [{\citenamefont {Syperek}\ \emph {et~al.}(2010)\citenamefont
  {Syperek}, \citenamefont {Leszczy{\'n}ski}, \citenamefont {Misiewicz},
  \citenamefont {Pavelescu}, \citenamefont {Gilfert},\ and\ \citenamefont
  {Reithmaier}}]{syperek2010time}%
  \BibitemOpen
  \bibfield  {author} {\bibinfo {author} {\bibfnamefont {M.}~\bibnamefont
  {Syperek}}, \bibinfo {author} {\bibfnamefont {P.}~\bibnamefont
  {Leszczy{\'n}ski}}, \bibinfo {author} {\bibfnamefont {J.}~\bibnamefont
  {Misiewicz}}, \bibinfo {author} {\bibfnamefont {E.~M.}\ \bibnamefont
  {Pavelescu}}, \bibinfo {author} {\bibfnamefont {C.}~\bibnamefont {Gilfert}},
  \ and\ \bibinfo {author} {\bibfnamefont {J.~P.}\ \bibnamefont {Reithmaier}},\
  }\href@noop {} {\bibfield  {journal} {\bibinfo  {journal} {Appl. Phys.
  Lett.}\ }\textbf {\bibinfo {volume} {96}},\ \bibinfo {pages} {011901}
  (\bibinfo {year} {2010})}\BibitemShut {NoStop}%
\bibitem [{\citenamefont {Syperek}\ \emph {et~al.}(2012)\citenamefont
  {Syperek}, \citenamefont {Andrzejewski}, \citenamefont {Rudno-Rudzi{\'n}ski},
  \citenamefont {S{\k{e}}k}, \citenamefont {Misiewicz}, \citenamefont
  {Pavelescu}, \citenamefont {Gilfert},\ and\ \citenamefont
  {Reithmaier}}]{syperek2012influence}%
  \BibitemOpen
  \bibfield  {author} {\bibinfo {author} {\bibfnamefont {M.}~\bibnamefont
  {Syperek}}, \bibinfo {author} {\bibfnamefont {J.}~\bibnamefont
  {Andrzejewski}}, \bibinfo {author} {\bibfnamefont {W.}~\bibnamefont
  {Rudno-Rudzi{\'n}ski}}, \bibinfo {author} {\bibfnamefont {G.}~\bibnamefont
  {S{\k{e}}k}}, \bibinfo {author} {\bibfnamefont {J.}~\bibnamefont
  {Misiewicz}}, \bibinfo {author} {\bibfnamefont {E.~M.}\ \bibnamefont
  {Pavelescu}}, \bibinfo {author} {\bibfnamefont {C.}~\bibnamefont {Gilfert}},
  \ and\ \bibinfo {author} {\bibfnamefont {J.~P.}\ \bibnamefont {Reithmaier}},\
  }\href@noop {} {\bibfield  {journal} {\bibinfo  {journal} {Phys. Rev. B}\
  }\textbf {\bibinfo {volume} {85}},\ \bibinfo {pages} {125311} (\bibinfo
  {year} {2012})}\BibitemShut {NoStop}%
\bibitem [{\citenamefont {Rudno-Rudzi{\'n}ski}\ \emph
  {et~al.}(2012)\citenamefont {Rudno-Rudzi{\'n}ski}, \citenamefont {S{\k{e}}k},
  \citenamefont {Andrzejewski}, \citenamefont {Misiewicz}, \citenamefont
  {Lelarge},\ and\ \citenamefont {Rousseau}}]{rudno2012electronic}%
  \BibitemOpen
  \bibfield  {author} {\bibinfo {author} {\bibfnamefont {W.}~\bibnamefont
  {Rudno-Rudzi{\'n}ski}}, \bibinfo {author} {\bibfnamefont {G.}~\bibnamefont
  {S{\k{e}}k}}, \bibinfo {author} {\bibfnamefont {J.}~\bibnamefont
  {Andrzejewski}}, \bibinfo {author} {\bibfnamefont {J.}~\bibnamefont
  {Misiewicz}}, \bibinfo {author} {\bibfnamefont {F.}~\bibnamefont {Lelarge}},
  \ and\ \bibinfo {author} {\bibfnamefont {B.}~\bibnamefont {Rousseau}},\
  }\href@noop {} {\bibfield  {journal} {\bibinfo  {journal} {Semiconductor
  Science and Technology}\ }\textbf {\bibinfo {volume} {27}},\ \bibinfo {pages}
  {105015} (\bibinfo {year} {2012})}\BibitemShut {NoStop}%
\bibitem [{\citenamefont {Inoshita}\ and\ \citenamefont
  {Sakaki}(1992)}]{Inoshita:92}%
  \BibitemOpen
  \bibfield  {author} {\bibinfo {author} {\bibfnamefont {T.}~\bibnamefont
  {Inoshita}}\ and\ \bibinfo {author} {\bibfnamefont {H.}~\bibnamefont
  {Sakaki}},\ }\href@noop {} {\bibfield  {journal} {\bibinfo  {journal} {Phys.
  Rev. B}\ }\textbf {\bibinfo {volume} {\textbf{46}}},\ \bibinfo {pages} {7260}
  (\bibinfo {year} {1992})}\BibitemShut {NoStop}%
\bibitem [{\citenamefont {Jiang}\ and\ \citenamefont {Singh}(1998)}]{Singh:98}%
  \BibitemOpen
  \bibfield  {author} {\bibinfo {author} {\bibfnamefont {H.}~\bibnamefont
  {Jiang}}\ and\ \bibinfo {author} {\bibfnamefont {J.}~\bibnamefont {Singh}},\
  }\href@noop {} {\bibfield  {journal} {\bibinfo  {journal} {IEEE J. Quantum
  Electron.}\ }\textbf {\bibinfo {volume} {\textbf{34}}},\ \bibinfo {pages}
  {1188} (\bibinfo {year} {1998})}\BibitemShut {NoStop}%
\bibitem [{\citenamefont {Seebeck}\ \emph {et~al.}(2005)\citenamefont
  {Seebeck}, \citenamefont {Nielsen}, \citenamefont {Gartner},\ and\
  \citenamefont {Jahnke}}]{Seebeck:05}%
  \BibitemOpen
  \bibfield  {author} {\bibinfo {author} {\bibfnamefont {J.}~\bibnamefont
  {Seebeck}}, \bibinfo {author} {\bibfnamefont {T.~R.}\ \bibnamefont
  {Nielsen}}, \bibinfo {author} {\bibfnamefont {P.}~\bibnamefont {Gartner}}, \
  and\ \bibinfo {author} {\bibfnamefont {F.}~\bibnamefont {Jahnke}},\
  }\href@noop {} {\bibfield  {journal} {\bibinfo  {journal} {Phys. Rev. B}\
  }\textbf {\bibinfo {volume} {\textbf{71}}},\ \bibinfo {pages} {125327}
  (\bibinfo {year} {2005})}\BibitemShut {NoStop}%
\bibitem [{\citenamefont {Bockelmann}\ and\ \citenamefont
  {Egeler}(1992)}]{Bockelmann:92}%
  \BibitemOpen
  \bibfield  {author} {\bibinfo {author} {\bibfnamefont {U.}~\bibnamefont
  {Bockelmann}}\ and\ \bibinfo {author} {\bibfnamefont {T.}~\bibnamefont
  {Egeler}},\ }\href@noop {} {\bibfield  {journal} {\bibinfo  {journal} {Phys.
  Rev. B}\ }\textbf {\bibinfo {volume} {\textbf{46}}},\ \bibinfo {pages}
  {15574} (\bibinfo {year} {1992})}\BibitemShut {NoStop}%
\bibitem [{\citenamefont {Uskov}\ \emph {et~al.}(1997)\citenamefont {Uskov},
  \citenamefont {Adler}, \citenamefont {Schweizer},\ and\ \citenamefont
  {Pilkuhn}}]{Uskov:97}%
  \BibitemOpen
  \bibfield  {author} {\bibinfo {author} {\bibfnamefont {A.~V.}\ \bibnamefont
  {Uskov}}, \bibinfo {author} {\bibfnamefont {F.}~\bibnamefont {Adler}},
  \bibinfo {author} {\bibfnamefont {H.}~\bibnamefont {Schweizer}}, \ and\
  \bibinfo {author} {\bibfnamefont {M.~H.}\ \bibnamefont {Pilkuhn}},\
  }\href@noop {} {\bibfield  {journal} {\bibinfo  {journal} {J. Appl. Phys.}\
  }\textbf {\bibinfo {volume} {\textbf{81}}},\ \bibinfo {pages} {7895}
  (\bibinfo {year} {1997})}\BibitemShut {NoStop}%
\bibitem [{\citenamefont {Magnusdottir}\ \emph {et~al.}(2003)\citenamefont
  {Magnusdottir}, \citenamefont {Bischoff}, \citenamefont {Uskov},\ and\
  \citenamefont {M{\o}rk}}]{Magnusdottir:03}%
  \BibitemOpen
  \bibfield  {author} {\bibinfo {author} {\bibfnamefont {I.}~\bibnamefont
  {Magnusdottir}}, \bibinfo {author} {\bibfnamefont {S.}~\bibnamefont
  {Bischoff}}, \bibinfo {author} {\bibfnamefont {A.~V.}\ \bibnamefont {Uskov}},
  \ and\ \bibinfo {author} {\bibfnamefont {J.}~\bibnamefont {M{\o}rk}},\
  }\href@noop {} {\bibfield  {journal} {\bibinfo  {journal} {Phys. Rev. B}\
  }\textbf {\bibinfo {volume} {\textbf{67}}},\ \bibinfo {pages} {205326}
  (\bibinfo {year} {2003})}\BibitemShut {NoStop}%
\bibitem [{\citenamefont {Nielsen}\ \emph {et~al.}(2004)\citenamefont
  {Nielsen}, \citenamefont {Gartner},\ and\ \citenamefont
  {Jahnke}}]{Nielsen:04}%
  \BibitemOpen
  \bibfield  {author} {\bibinfo {author} {\bibfnamefont {T.~R.}\ \bibnamefont
  {Nielsen}}, \bibinfo {author} {\bibfnamefont {P.}~\bibnamefont {Gartner}}, \
  and\ \bibinfo {author} {\bibfnamefont {F.}~\bibnamefont {Jahnke}},\
  }\href@noop {} {\bibfield  {journal} {\bibinfo  {journal} {Phys. Rev. B}\
  }\textbf {\bibinfo {volume} {\textbf{69}}},\ \bibinfo {pages} {235314}
  (\bibinfo {year} {2004})}\BibitemShut {NoStop}%
\bibitem [{\citenamefont {Pavelescu}\ \emph {et~al.}(2009)\citenamefont
  {Pavelescu}, \citenamefont {Gilfert}, \citenamefont {Reithmaier},
  \citenamefont {Martin-Minguez},\ and\ \citenamefont
  {Esquivias}}]{pavelescu2009high}%
  \BibitemOpen
  \bibfield  {author} {\bibinfo {author} {\bibfnamefont {E.-M.}\ \bibnamefont
  {Pavelescu}}, \bibinfo {author} {\bibfnamefont {C.}~\bibnamefont {Gilfert}},
  \bibinfo {author} {\bibfnamefont {J.~P.}\ \bibnamefont {Reithmaier}},
  \bibinfo {author} {\bibfnamefont {A.}~\bibnamefont {Martin-Minguez}}, \ and\
  \bibinfo {author} {\bibfnamefont {I.}~\bibnamefont {Esquivias}},\ }\href@noop
  {} {\bibfield  {journal} {\bibinfo  {journal} {IEEE Photonics Technology
  Letters}\ }\textbf {\bibinfo {volume} {21}},\ \bibinfo {pages} {999}
  (\bibinfo {year} {2009})}\BibitemShut {NoStop}%
\bibitem [{\citenamefont {Pulka}\ \emph {et~al.}(2012)\citenamefont {Pulka},
  \citenamefont {Piwonski}, \citenamefont {Huyet}, \citenamefont {Houlihan},
  \citenamefont {Semenova}, \citenamefont {Lematre}, \citenamefont {Merghem},
  \citenamefont {Martinez},\ and\ \citenamefont
  {Ramdane}}]{pulka2012ultrafast}%
  \BibitemOpen
  \bibfield  {author} {\bibinfo {author} {\bibfnamefont {J.}~\bibnamefont
  {Pulka}}, \bibinfo {author} {\bibfnamefont {T.}~\bibnamefont {Piwonski}},
  \bibinfo {author} {\bibfnamefont {G.}~\bibnamefont {Huyet}}, \bibinfo
  {author} {\bibfnamefont {J.}~\bibnamefont {Houlihan}}, \bibinfo {author}
  {\bibfnamefont {E.}~\bibnamefont {Semenova}}, \bibinfo {author}
  {\bibfnamefont {A.}~\bibnamefont {Lematre}}, \bibinfo {author} {\bibfnamefont
  {K.}~\bibnamefont {Merghem}}, \bibinfo {author} {\bibfnamefont
  {A.}~\bibnamefont {Martinez}}, \ and\ \bibinfo {author} {\bibfnamefont
  {A.}~\bibnamefont {Ramdane}},\ }\href@noop {} {\bibfield  {journal} {\bibinfo
   {journal} {Appl. Phys. Lett.}\ }\textbf {\bibinfo {volume} {100}},\ \bibinfo
  {pages} {071107} (\bibinfo {year} {2012})}\BibitemShut {NoStop}%
\bibitem [{\citenamefont {Mary{\'n}ski}\ \emph {et~al.}(2013)\citenamefont
  {Mary{\'n}ski}, \citenamefont {S{\k{e}}k}, \citenamefont {Musia{\l}},
  \citenamefont {Andrzejewski}, \citenamefont {Misiewicz}, \citenamefont
  {Gilfert}, \citenamefont {Reithmaier}, \citenamefont {Capua}, \citenamefont
  {Karni}, \citenamefont {Gready} \emph {et~al.}}]{marynski2013electronic}%
  \BibitemOpen
  \bibfield  {author} {\bibinfo {author} {\bibfnamefont {A.}~\bibnamefont
  {Mary{\'n}ski}}, \bibinfo {author} {\bibfnamefont {G.}~\bibnamefont
  {S{\k{e}}k}}, \bibinfo {author} {\bibfnamefont {A.}~\bibnamefont
  {Musia{\l}}}, \bibinfo {author} {\bibfnamefont {J.}~\bibnamefont
  {Andrzejewski}}, \bibinfo {author} {\bibfnamefont {J.}~\bibnamefont
  {Misiewicz}}, \bibinfo {author} {\bibfnamefont {C.}~\bibnamefont {Gilfert}},
  \bibinfo {author} {\bibfnamefont {J.~P.}\ \bibnamefont {Reithmaier}},
  \bibinfo {author} {\bibfnamefont {A.}~\bibnamefont {Capua}}, \bibinfo
  {author} {\bibfnamefont {O.}~\bibnamefont {Karni}}, \bibinfo {author}
  {\bibfnamefont {D.}~\bibnamefont {Gready}},  \emph {et~al.},\ }\href@noop {}
  {\bibfield  {journal} {\bibinfo  {journal} {Journal of Applied Physics}\
  }\textbf {\bibinfo {volume} {114}},\ \bibinfo {pages} {094306} (\bibinfo
  {year} {2013})}\BibitemShut {NoStop}%
\bibitem [{\citenamefont {Banyoudeh}\ and\ \citenamefont
  {Reithmaier}(2015)}]{banyoudeh2015high}%
  \BibitemOpen
  \bibfield  {author} {\bibinfo {author} {\bibfnamefont {S.}~\bibnamefont
  {Banyoudeh}}\ and\ \bibinfo {author} {\bibfnamefont {J.~P.}\ \bibnamefont
  {Reithmaier}},\ }\href@noop {} {\bibfield  {journal} {\bibinfo  {journal}
  {Journal of Crystal Growth}\ }\textbf {\bibinfo {volume} {425}},\ \bibinfo
  {pages} {299} (\bibinfo {year} {2015})}\BibitemShut {NoStop}%
\bibitem [{\citenamefont {Syperek}\ \emph {et~al.}(2016)\citenamefont
  {Syperek}, \citenamefont {Dusanowski}, \citenamefont {Gawe{\l}czyk},
  \citenamefont {S{\k{e}}k}, \citenamefont {Somers}, \citenamefont
  {Reithmaier}, \citenamefont {H{\"o}fling},\ and\ \citenamefont
  {Misiewicz}}]{syperek2016exciton}%
  \BibitemOpen
  \bibfield  {author} {\bibinfo {author} {\bibfnamefont {M.}~\bibnamefont
  {Syperek}}, \bibinfo {author} {\bibfnamefont {{\L}.}~\bibnamefont
  {Dusanowski}}, \bibinfo {author} {\bibfnamefont {M.}~\bibnamefont
  {Gawe{\l}czyk}}, \bibinfo {author} {\bibfnamefont {G.}~\bibnamefont
  {S{\k{e}}k}}, \bibinfo {author} {\bibfnamefont {A.}~\bibnamefont {Somers}},
  \bibinfo {author} {\bibfnamefont {J.~P.}\ \bibnamefont {Reithmaier}},
  \bibinfo {author} {\bibfnamefont {S.}~\bibnamefont {H{\"o}fling}}, \ and\
  \bibinfo {author} {\bibfnamefont {J.}~\bibnamefont {Misiewicz}},\ }\href@noop
  {} {\bibfield  {journal} {\bibinfo  {journal} {Appl. Phys. Lett.}\ }\textbf
  {\bibinfo {volume} {109}},\ \bibinfo {pages} {193108} (\bibinfo {year}
  {2016})}\BibitemShut {NoStop}%
\bibitem [{\citenamefont {Bauer}\ \emph {et~al.}(2018)\citenamefont {Bauer},
  \citenamefont {Sichkovskyi},\ and\ \citenamefont
  {Reithmaier}}]{bauer2018growth}%
  \BibitemOpen
  \bibfield  {author} {\bibinfo {author} {\bibfnamefont {S.}~\bibnamefont
  {Bauer}}, \bibinfo {author} {\bibfnamefont {V.}~\bibnamefont {Sichkovskyi}},
  \ and\ \bibinfo {author} {\bibfnamefont {J.~P.}\ \bibnamefont {Reithmaier}},\
  }\href@noop {} {\bibfield  {journal} {\bibinfo  {journal} {Journal of Crystal
  Growth}\ }\textbf {\bibinfo {volume} {491}},\ \bibinfo {pages} {20} (\bibinfo
  {year} {2018})}\BibitemShut {NoStop}%
\bibitem [{\citenamefont {Rudno-Rudzi{\'n}ski}\ \emph
  {et~al.}(2018)\citenamefont {Rudno-Rudzi{\'n}ski}, \citenamefont
  {Biega{\'n}ska}, \citenamefont {Misiewicz}, \citenamefont {Lelarge},
  \citenamefont {Rousseau},\ and\ \citenamefont
  {S{\k{e}}k}}]{rudno2018carrier}%
  \BibitemOpen
  \bibfield  {author} {\bibinfo {author} {\bibfnamefont {W.}~\bibnamefont
  {Rudno-Rudzi{\'n}ski}}, \bibinfo {author} {\bibfnamefont {D.}~\bibnamefont
  {Biega{\'n}ska}}, \bibinfo {author} {\bibfnamefont {J.}~\bibnamefont
  {Misiewicz}}, \bibinfo {author} {\bibfnamefont {F.}~\bibnamefont {Lelarge}},
  \bibinfo {author} {\bibfnamefont {B.}~\bibnamefont {Rousseau}}, \ and\
  \bibinfo {author} {\bibfnamefont {G.}~\bibnamefont {S{\k{e}}k}},\ }\href@noop
  {} {\bibfield  {journal} {\bibinfo  {journal} {Applied Physics Letters}\
  }\textbf {\bibinfo {volume} {112}},\ \bibinfo {pages} {051103} (\bibinfo
  {year} {2018})}\BibitemShut {NoStop}%
\bibitem [{\citenamefont {Rudno-Rudzi{\'n}ski}\ \emph
  {et~al.}(2017)\citenamefont {Rudno-Rudzi{\'n}ski}, \citenamefont {Syperek},
  \citenamefont {Mary{\'n}ski}, \citenamefont {Andrzejewski}, \citenamefont
  {Misiewicz}, \citenamefont {Bauer}, \citenamefont {Sichkovskyi},
  \citenamefont {Reithmaier}, \citenamefont {Schowalter}, \citenamefont
  {Gerken} \emph {et~al.}}]{rudno2018control}%
  \BibitemOpen
  \bibfield  {author} {\bibinfo {author} {\bibfnamefont {W.}~\bibnamefont
  {Rudno-Rudzi{\'n}ski}}, \bibinfo {author} {\bibfnamefont {M.}~\bibnamefont
  {Syperek}}, \bibinfo {author} {\bibfnamefont {A.}~\bibnamefont
  {Mary{\'n}ski}}, \bibinfo {author} {\bibfnamefont {J.}~\bibnamefont
  {Andrzejewski}}, \bibinfo {author} {\bibfnamefont {J.}~\bibnamefont
  {Misiewicz}}, \bibinfo {author} {\bibfnamefont {S.}~\bibnamefont {Bauer}},
  \bibinfo {author} {\bibfnamefont {V.~I.}\ \bibnamefont {Sichkovskyi}},
  \bibinfo {author} {\bibfnamefont {J.~P.}\ \bibnamefont {Reithmaier}},
  \bibinfo {author} {\bibfnamefont {M.}~\bibnamefont {Schowalter}}, \bibinfo
  {author} {\bibfnamefont {B.}~\bibnamefont {Gerken}},  \emph {et~al.},\
  }\href@noop {} {\bibfield  {journal} {\bibinfo  {journal} {physica status
  solidi (a)}\ }\textbf {\bibinfo {volume} {215}},\ \bibinfo {pages} {1700455}
  (\bibinfo {year} {2017})}\BibitemShut {NoStop}%
\bibitem [{\citenamefont {Syperek}\ \emph {et~al.}(2018)\citenamefont
  {Syperek}, \citenamefont {Andrzejewski}, \citenamefont {Rogowicz},
  \citenamefont {Misiewicz}, \citenamefont {Bauer}, \citenamefont
  {Sichkovskyi}, \citenamefont {Reithmaier},\ and\ \citenamefont
  {S{\k{e}}k}}]{syperek2018carrier}%
  \BibitemOpen
  \bibfield  {author} {\bibinfo {author} {\bibfnamefont {M.}~\bibnamefont
  {Syperek}}, \bibinfo {author} {\bibfnamefont {J.}~\bibnamefont
  {Andrzejewski}}, \bibinfo {author} {\bibfnamefont {E.}~\bibnamefont
  {Rogowicz}}, \bibinfo {author} {\bibfnamefont {J.}~\bibnamefont {Misiewicz}},
  \bibinfo {author} {\bibfnamefont {S.}~\bibnamefont {Bauer}}, \bibinfo
  {author} {\bibfnamefont {V.}~\bibnamefont {Sichkovskyi}}, \bibinfo {author}
  {\bibfnamefont {J.}~\bibnamefont {Reithmaier}}, \ and\ \bibinfo {author}
  {\bibfnamefont {G.}~\bibnamefont {S{\k{e}}k}},\ }\href@noop {} {\bibfield
  {journal} {\bibinfo  {journal} {Applied Physics Letters}\ }\textbf {\bibinfo
  {volume} {112}},\ \bibinfo {pages} {221901} (\bibinfo {year}
  {2018})}\BibitemShut {NoStop}%
\bibitem [{\citenamefont {Bhattacharya}\ \emph {et~al.}(1996)\citenamefont
  {Bhattacharya}, \citenamefont {Singh}, \citenamefont {Yoon}, \citenamefont
  {Zhang}, \citenamefont {Gutierrez-Aitken},\ and\ \citenamefont
  {Lam}}]{bhattacharya1996tunneling}%
  \BibitemOpen
  \bibfield  {author} {\bibinfo {author} {\bibfnamefont {P.}~\bibnamefont
  {Bhattacharya}}, \bibinfo {author} {\bibfnamefont {J.}~\bibnamefont {Singh}},
  \bibinfo {author} {\bibfnamefont {H.}~\bibnamefont {Yoon}}, \bibinfo {author}
  {\bibfnamefont {X.}~\bibnamefont {Zhang}}, \bibinfo {author} {\bibfnamefont
  {A.}~\bibnamefont {Gutierrez-Aitken}}, \ and\ \bibinfo {author}
  {\bibfnamefont {Y.}~\bibnamefont {Lam}},\ }\href@noop {} {\bibfield
  {journal} {\bibinfo  {journal} {IEEE J. Quantum Electron.}\ }\textbf
  {\bibinfo {volume} {32}},\ \bibinfo {pages} {1620} (\bibinfo {year}
  {1996})}\BibitemShut {NoStop}%
\bibitem [{\citenamefont {Asryan}\ and\ \citenamefont
  {Luryi}(2001)}]{asryan2001tunneling}%
  \BibitemOpen
  \bibfield  {author} {\bibinfo {author} {\bibfnamefont {L.~V.}\ \bibnamefont
  {Asryan}}\ and\ \bibinfo {author} {\bibfnamefont {S.}~\bibnamefont {Luryi}},\
  }\href@noop {} {\bibfield  {journal} {\bibinfo  {journal} {IEEE J. Quantum
  Electron.}\ }\textbf {\bibinfo {volume} {37}},\ \bibinfo {pages} {905}
  (\bibinfo {year} {2001})}\BibitemShut {NoStop}%
\bibitem [{\citenamefont {Han}\ and\ \citenamefont
  {Asryan}(2008)}]{han2008tunneling}%
  \BibitemOpen
  \bibfield  {author} {\bibinfo {author} {\bibfnamefont {D.-S.}\ \bibnamefont
  {Han}}\ and\ \bibinfo {author} {\bibfnamefont {L.~V.}\ \bibnamefont
  {Asryan}},\ }\href@noop {} {\bibfield  {journal} {\bibinfo  {journal} {Appl.
  Phys. Lett.}\ }\textbf {\bibinfo {volume} {92}},\ \bibinfo {pages} {251113}
  (\bibinfo {year} {2008})}\BibitemShut {NoStop}%
\bibitem [{\citenamefont {Gready}\ and\ \citenamefont
  {Eisenstein}(2010)}]{gready2010carrier}%
  \BibitemOpen
  \bibfield  {author} {\bibinfo {author} {\bibfnamefont {D.}~\bibnamefont
  {Gready}}\ and\ \bibinfo {author} {\bibfnamefont {G.}~\bibnamefont
  {Eisenstein}},\ }\href@noop {} {\bibfield  {journal} {\bibinfo  {journal}
  {IEEE J. Quantum Electron.}\ }\textbf {\bibinfo {volume} {46}},\ \bibinfo
  {pages} {1611} (\bibinfo {year} {2010})}\BibitemShut {NoStop}%
\bibitem [{\citenamefont {Gready}\ and\ \citenamefont
  {Eisenstein}(2011)}]{gready2011effects}%
  \BibitemOpen
  \bibfield  {author} {\bibinfo {author} {\bibfnamefont {D.}~\bibnamefont
  {Gready}}\ and\ \bibinfo {author} {\bibfnamefont {G.}~\bibnamefont
  {Eisenstein}},\ }\href@noop {} {\bibfield  {journal} {\bibinfo  {journal}
  {IEEE J. Quantum Electron.}\ }\textbf {\bibinfo {volume} {47}},\ \bibinfo
  {pages} {944} (\bibinfo {year} {2011})}\BibitemShut {NoStop}%
\bibitem [{\citenamefont {Chang}\ \emph {et~al.}(2004)\citenamefont {Chang},
  \citenamefont {Chuang},\ and\ \citenamefont {Holonyak~Jr}}]{chang2004phonon}%
  \BibitemOpen
  \bibfield  {author} {\bibinfo {author} {\bibfnamefont {S.-W.}\ \bibnamefont
  {Chang}}, \bibinfo {author} {\bibfnamefont {S.-L.}\ \bibnamefont {Chuang}}, \
  and\ \bibinfo {author} {\bibfnamefont {N.}~\bibnamefont {Holonyak~Jr}},\
  }\href@noop {} {\bibfield  {journal} {\bibinfo  {journal} {Phys. Rev. B}\
  }\textbf {\bibinfo {volume} {70}},\ \bibinfo {pages} {125312} (\bibinfo
  {year} {2004})}\BibitemShut {NoStop}%
\bibitem [{\citenamefont {Mielnik-Pyszczorski}\ \emph
  {et~al.}(2015)\citenamefont {Mielnik-Pyszczorski}, \citenamefont
  {Gawarecki},\ and\ \citenamefont {Machnikowski}}]{mielnik2015phonon}%
  \BibitemOpen
  \bibfield  {author} {\bibinfo {author} {\bibfnamefont {A.}~\bibnamefont
  {Mielnik-Pyszczorski}}, \bibinfo {author} {\bibfnamefont {K.}~\bibnamefont
  {Gawarecki}}, \ and\ \bibinfo {author} {\bibfnamefont {P.}~\bibnamefont
  {Machnikowski}},\ }\href@noop {} {\bibfield  {journal} {\bibinfo  {journal}
  {Phys. Rev. B}\ }\textbf {\bibinfo {volume} {91}},\ \bibinfo {pages} {195421}
  (\bibinfo {year} {2015})}\BibitemShut {NoStop}%
\bibitem [{\citenamefont {Popescu}\ \emph {et~al.}(2009)\citenamefont
  {Popescu}, \citenamefont {Bester},\ and\ \citenamefont
  {Zunger}}]{PhysRevB.80.045327}%
  \BibitemOpen
  \bibfield  {author} {\bibinfo {author} {\bibfnamefont {V.}~\bibnamefont
  {Popescu}}, \bibinfo {author} {\bibfnamefont {G.}~\bibnamefont {Bester}}, \
  and\ \bibinfo {author} {\bibfnamefont {A.}~\bibnamefont {Zunger}},\ }\href
  {\doibase 10.1103/PhysRevB.80.045327} {\bibfield  {journal} {\bibinfo
  {journal} {Phys. Rev. B}\ }\textbf {\bibinfo {volume} {80}},\ \bibinfo
  {pages} {045327} (\bibinfo {year} {2009})}\BibitemShut {NoStop}%
\bibitem [{\citenamefont {Hackenbuchner}(2002)}]{nextnano3}%
  \BibitemOpen
  \bibfield  {author} {\bibinfo {author} {\bibfnamefont {S.}~\bibnamefont
  {Hackenbuchner}},\ }\href@noop {} {\emph {\bibinfo {title} {'Elektronische
  Struktur von Halbleiter-Nanobauelementen im thermodynamischen
  Nichtgleichgewicht' in Selected Topics of Semiconductor Physics and
  Technology}}},\ edited by\ \bibinfo {editor} {\bibfnamefont {G.}~\bibnamefont
  {Abstreiter}}, \bibinfo {editor} {\bibfnamefont {M.-C.}\ \bibnamefont
  {Amann}}, \bibinfo {editor} {\bibfnamefont {M.}~\bibnamefont {Stutzmann}}, \
  and\ \bibinfo {editor} {\bibfnamefont {P.}~\bibnamefont {Vogl}},\
  Vol.~\bibinfo {volume} {48}\ (\bibinfo  {publisher} {Verein zur F{\"o}rderung
  des Walter Schottky Instituts der Technischen Universit{\"a}t M{\"u}nchen
  e.V.},\ \bibinfo {address} {M{\"u}nchen},\ \bibinfo {year}
  {2002})\BibitemShut {NoStop}%
\bibitem [{\citenamefont {Callaway}(1991)}]{CALLAWAY:1991611}%
  \BibitemOpen
  \bibfield  {author} {\bibinfo {author} {\bibfnamefont {J.}~\bibnamefont
  {Callaway}},\ }\href {\doibase
  https://doi.org/10.1016/B978-0-12-155203-9.50012-2} {\emph {\bibinfo {title}
  {Quantum Theory of the Solid State}}},\ \bibinfo {edition} {2nd}\ ed.\
  (\bibinfo  {publisher} {Academic Press},\ \bibinfo {address} {Boston},\
  \bibinfo {year} {1991})\ pp.\ \bibinfo {pages} {611 -- 760}\BibitemShut
  {NoStop}%
\bibitem [{\citenamefont {Inoshita}\ and\ \citenamefont
  {Sakaki}(1997)}]{Inoshita:97}%
  \BibitemOpen
  \bibfield  {author} {\bibinfo {author} {\bibfnamefont {T.}~\bibnamefont
  {Inoshita}}\ and\ \bibinfo {author} {\bibfnamefont {H.}~\bibnamefont
  {Sakaki}},\ }\href@noop {} {\bibfield  {journal} {\bibinfo  {journal} {Phys.
  Rev. B}\ }\textbf {\bibinfo {volume} {\textbf{56}}},\ \bibinfo {pages}
  {R4355} (\bibinfo {year} {1997})}\BibitemShut {NoStop}%
\bibitem [{\citenamefont {Kr{\'a}l}\ and\ \citenamefont
  {Kh{\'a}s}(1998)}]{Kral:98}%
  \BibitemOpen
  \bibfield  {author} {\bibinfo {author} {\bibfnamefont {K.}~\bibnamefont
  {Kr{\'a}l}}\ and\ \bibinfo {author} {\bibfnamefont {Z.}~\bibnamefont
  {Kh{\'a}s}},\ }\href@noop {} {\bibfield  {journal} {\bibinfo  {journal}
  {Phys. Rev. B}\ }\textbf {\bibinfo {volume} {\textbf{57}}},\ \bibinfo {pages}
  {R2061} (\bibinfo {year} {1998})}\BibitemShut {NoStop}%
\bibitem [{\citenamefont {Haug}\ and\ \citenamefont
  {Koch}(2004)}]{Haug_Koch:04}%
  \BibitemOpen
  \bibfield  {author} {\bibinfo {author} {\bibfnamefont {H.}~\bibnamefont
  {Haug}}\ and\ \bibinfo {author} {\bibfnamefont {S.}~\bibnamefont {Koch}},\
  }\href@noop {} {\emph {\bibinfo {title} {Quantum Theory of the Optical and
  Electronic Properties of Semiconductors}}},\ \bibinfo {edition} {4th}\ ed.\
  (\bibinfo  {publisher} {World Scientific Publ.},\ \bibinfo {address}
  {Singapore},\ \bibinfo {year} {2004})\BibitemShut {NoStop}%
\bibitem [{\citenamefont {B{\'a}nyai}\ \emph {et~al.}(1998)\citenamefont
  {B{\'a}nyai}, \citenamefont {Gartner},\ and\ \citenamefont
  {Haug}}]{Banyai:98}%
  \BibitemOpen
  \bibfield  {author} {\bibinfo {author} {\bibfnamefont {L.}~\bibnamefont
  {B{\'a}nyai}}, \bibinfo {author} {\bibfnamefont {P.}~\bibnamefont {Gartner}},
  \ and\ \bibinfo {author} {\bibfnamefont {H.}~\bibnamefont {Haug}},\
  }\href@noop {} {\bibfield  {journal} {\bibinfo  {journal} {Eur. Phys. J. B}\
  }\textbf {\bibinfo {volume} {\textbf{1}}},\ \bibinfo {pages} {209} (\bibinfo
  {year} {1998})}\BibitemShut {NoStop}%
\bibitem [{\citenamefont {Gartner}\ \emph {et~al.}(2002)\citenamefont
  {Gartner}, \citenamefont {B{\'a}nyai},\ and\ \citenamefont
  {Haug}}]{Gartner:02}%
  \BibitemOpen
  \bibfield  {author} {\bibinfo {author} {\bibfnamefont {P.}~\bibnamefont
  {Gartner}}, \bibinfo {author} {\bibfnamefont {L.}~\bibnamefont {B{\'a}nyai}},
  \ and\ \bibinfo {author} {\bibfnamefont {H.}~\bibnamefont {Haug}},\
  }\href@noop {} {\bibfield  {journal} {\bibinfo  {journal} {Phys. Rev. B}\
  }\textbf {\bibinfo {volume} {\textbf{66}}},\ \bibinfo {pages} {75205}
  (\bibinfo {year} {2002})}\BibitemShut {NoStop}%
\bibitem [{\citenamefont {Lorke}\ \emph
  {et~al.}(2006{\natexlab{a}})\citenamefont {Lorke}, \citenamefont {Nielsen},
  \citenamefont {Seebeck}, \citenamefont {Gartner},\ and\ \citenamefont
  {Jahnke}}]{Lorke:06c}%
  \BibitemOpen
  \bibfield  {author} {\bibinfo {author} {\bibfnamefont {M.}~\bibnamefont
  {Lorke}}, \bibinfo {author} {\bibfnamefont {T.~R.}\ \bibnamefont {Nielsen}},
  \bibinfo {author} {\bibfnamefont {J.}~\bibnamefont {Seebeck}}, \bibinfo
  {author} {\bibfnamefont {P.}~\bibnamefont {Gartner}}, \ and\ \bibinfo
  {author} {\bibfnamefont {F.}~\bibnamefont {Jahnke}},\ }\href
  {http://stacks.iop.org/1742-6596/35/i=1/a=016} {\bibfield  {journal}
  {\bibinfo  {journal} {Journal of Physics: Conference Series}\ }\textbf
  {\bibinfo {volume} {35}},\ \bibinfo {pages} {182} (\bibinfo {year}
  {2006}{\natexlab{a}})}\BibitemShut {NoStop}%
\bibitem [{\citenamefont {Lorke}\ \emph {et~al.}(2011)\citenamefont {Lorke},
  \citenamefont {Nielsen},\ and\ \citenamefont {M{\o}rk}}]{Lorke:11}%
  \BibitemOpen
  \bibfield  {author} {\bibinfo {author} {\bibfnamefont {M.}~\bibnamefont
  {Lorke}}, \bibinfo {author} {\bibfnamefont {T.~R.}\ \bibnamefont {Nielsen}},
  \ and\ \bibinfo {author} {\bibfnamefont {J.}~\bibnamefont {M{\o}rk}},\ }\href
  {\doibase DOI:10.1063/1.3651765} {\bibfield  {journal} {\bibinfo  {journal}
  {Appl. Phys. Lett.}\ }\textbf {\bibinfo {volume} {99}},\ \bibinfo {pages}
  {151110} (\bibinfo {year} {2011})}\BibitemShut {NoStop}%
\bibitem [{\citenamefont {Lorke}\ \emph
  {et~al.}(2006{\natexlab{b}})\citenamefont {Lorke}, \citenamefont {Nielsen},
  \citenamefont {Seebeck}, \citenamefont {Gartner},\ and\ \citenamefont
  {Jahnke}}]{Lorke:06}%
  \BibitemOpen
  \bibfield  {author} {\bibinfo {author} {\bibfnamefont {M.}~\bibnamefont
  {Lorke}}, \bibinfo {author} {\bibfnamefont {T.~R.}\ \bibnamefont {Nielsen}},
  \bibinfo {author} {\bibfnamefont {J.}~\bibnamefont {Seebeck}}, \bibinfo
  {author} {\bibfnamefont {P.}~\bibnamefont {Gartner}}, \ and\ \bibinfo
  {author} {\bibfnamefont {F.}~\bibnamefont {Jahnke}},\ }\href@noop {}
  {\bibfield  {journal} {\bibinfo  {journal} {Phys. Rev. B}\ }\textbf {\bibinfo
  {volume} {\textbf{73}}},\ \bibinfo {pages} {085324} (\bibinfo {year}
  {2006}{\natexlab{b}})}\BibitemShut {NoStop}%
\bibitem [{\citenamefont {Mahan}(1990)}]{Mahan:90}%
  \BibitemOpen
  \bibfield  {author} {\bibinfo {author} {\bibfnamefont {G.}~\bibnamefont
  {Mahan}},\ }\href@noop {} {\emph {\bibinfo {title} {Many-Particle Physics}}}\
  (\bibinfo  {publisher} {Plenum Press},\ \bibinfo {address} {New York},\
  \bibinfo {year} {1990})\BibitemShut {NoStop}%
\end{thebibliography}%

\end{document}